\documentclass[journal]{IEEEtran}
\ifCLASSINFOpdf
\else
\fi
\usepackage{amsmath,amsthm}
\usepackage{bm}
\usepackage{bbm}
\usepackage{float}
\usepackage{lipsum}
\usepackage{graphicx}
\newtheorem{Definition}{Definition}

\usepackage{booktabs} 
\usepackage{amssymb}
\usepackage{subfigure}
\usepackage{multirow}
\usepackage{svg}
\usepackage{verbatim}
\usepackage{siunitx}
\begin{document}
\title{Platoon-Centric Green Light Optimal Speed Advisory Using Safe Reinforcement Learning}

\author{Ruining Yang, Jingyuan Zhou, Qiqing Wang, Jinhao Liang, and Kaidi Yang
\thanks{Copyright (c) 2025 IEEE. Personal use of this material is permitted. However, permission to use this material for any other purposes must be obtained from the IEEE by sending a request to pubs-permissions@ieee.org.}
 \thanks{This research was supported by the Singapore Ministry of Education (MOE) under AcRF Tier 1 (A-8001183-00-00). This article
solely reflects the opinions and conclusions of its authors and not Singapore MOE or any other entity. (\emph{Corresponding Author: Kaidi Yang})}
\thanks{R. Yang is with the School of Civil and Environmental Engineering, Georgia Institute of Technology, GA 30318, USA. (Email: yruining@gatech.edu)}
\thanks{J. Zhou, Q. Wang, J. Liang, and K. Yang are with the Department of Civil and Environmental Engineering, National University of Singapore, Singapore 119077. (Email:{\{jingyuanzhou, qiqing.wang\}@u.nus.edu;\{jh.liang, kaidi.yang\}@nus.edu.sg}.)}
}

\markboth{IEEE Transactions on Vehicular Technology}
{Shell \MakeLowercase{\textit{et al.}}: Bare Demo of IEEEtran.cls for IEEE Journals}

\maketitle
\pagestyle{empty} 
\thispagestyle{empty}
\begin{abstract}
With recent advancements in Connected Autonomous Vehicles (CAVs), Green Light Optimal Speed Advisory (GLOSA) emerges as a promising eco-driving strategy to reduce the number of stops and idle time at intersections, thereby reducing energy consumption and emissions. Existing studies typically improve energy and travel efficiency for individual CAVs without considering their impacts on the entire mixed-traffic platoon, leading to inefficient traffic flow. While Reinforcement Learning (RL) has the potential to achieve platoon-level control in a mixed-traffic environment, the training of RL is still challenged by (i) car-following safety, i.e., CAVs should not collide with their immediate preceding vehicles, and (ii) red-light safety, i.e., CAVs should not run red lights.
To address these challenges, this paper develops a platoon-centric, safe RL-based GLOSA system that uses a multi-agent controller to optimize CAV speed while achieving a balance between energy consumption and travel efficiency.
We further incorporate Control Barrier Functions (CBFs) into the RL-based policy to provide explicit safety guarantees in terms of car-following safety and red-light safety. Our simulation results illustrate that our proposed method outperforms state-of-the-art methods in terms of driving safety and platoon energy consumption. 
\end{abstract}

\IEEEpeerreviewmaketitle

\section{Introduction} \label{sec:intro}
\IEEEPARstart{R}{oad} transportation consumes a large amount of fossil fuels, accounting for approximately $16\%$ of global energy consumption and $10\%$ of global greenhouse gas emissions \cite{IEA-report}.
To address these environmental challenges, various strategies have been proposed to reduce energy consumption and emissions, such as promoting Electric Vehicles (EVs) to mitigate the need for fossil fuels \cite{mohammadi2023towards, gnanavendan2024challenges}, developing efficient public transportation systems to reduce the reliance on private vehicles \cite{oeschger2020micromobility, kuo2023public}, and implementing eco-driving strategies that allow vehicles to adopt fuel-efficient driving practices (e.g., avoiding sudden accelerations) \cite{wu2023deep, yao2023optimal}. 
As a powerful technique in eco-driving, Green Light Optimal Speed Advisory (GLOSA) has attracted significant research attention, especially with advancements in Connected Autonomous Vehicles (CAVs)~\cite{chen2024review, cheng2024towards}. 
GLOSA systems are primarily implemented in urban areas with dense junctions, whereby 
CAVs receive signal timing information via Infrastructure-to-Vehicle (I2V) communication and accordingly perform vehicle trajectory planning to ensure that they arrive at the intersection during the green phase. Consequently, GLOSA systems allow CAVs to avoid frequent stopping and reduce delays at intersections, which further results in lower energy consumption and pollutant emissions. 

Generally speaking, existing methods for GLOSA systems can be divided into two categories: optimization-based methods and learning-based methods. Optimization-based methods build on a system dynamics model and seek to optimize certain energy-related criteria. For example, Ref. \cite{zhang2023platoon} developed a hybrid model predictive control method for a mixed-traffic platoon comprising CAVs and Human-driven Vehicles (HDVs) to optimize their speed profiles and platoon splitting decisions. Ref. \cite{typaldos2023modified} designed GLOSA systems with uncertain signal switching times using Differential Dynamic Programming (DDP). Despite the potential of these strategies to save energy, their performance heavily depends on the embedded system model, which needs to be sufficiently simple to ensure tractability yet accurate to ensure control performance. Nevertheless, GLOSA systems can be complex to model due to the nonlinear and heterogeneous behavior of the HDVs involved, making it challenging to trade off tractability and control performance for optimization-based methods.  

In contrast, learning-based methods, e.g., Reinforcement Learning (RL), learn a policy from interactions with the traffic systems to determine the GLOSA decisions based on real-time observations. For example, Ref. \cite{zhou2019development} leveraged Deep Deterministic Policy Gradient (DDPG) to reduce speed oscillations triggered by human drivers at signalized intersections. Ref. \cite{xu2023adaptive} developed a hybrid actor-critic RL-based GLOSA framework that enables the speed of CAV to change at an adaptive frequency. Ref. \cite{wegener2021automated} leveraged the twin-delayed deep deterministic policy gradient (TD3) method to automatically control vehicles using minimal sensor data and traffic light timings. Ref. \cite{zhu2022safe} proposed a safe model-based off-policy RL algorithm that integrates a trajectory optimizer to improve sample efficiency and a learned safe set to ensure driving safety.
These learning-based methods are promising, as they do not require a system model and are highly computationally efficient for real-time control. 

Despite their potential, existing learning-based methods still suffer from two limitations.  
First, most existing studies are vehicle-centric rather than platoon-centric, i.e., they focus on individual CAVs without considering the impacts of their actions on the following vehicles within a platoon. 
Compared to vehicle-centric control, the platoon-centric counterpart exhibits a better energy-saving potential thanks to its capability to jointly optimize the energy consumption of the controlled CAVs and their following vehicles.  
Nevertheless, platoon-centric control requires the characterization of the behavior of the following vehicles, which can be challenging in mixed-traffic scenarios where the information regarding some following HDVs may not be observable due to their lack of communication capabilities. To the best of our knowledge, few existing works have considered the platoon-centric GLOSA systems in mixed traffic. 
Second, existing RL-based approaches tend to consider safety with soft constraints in the reward function, which cannot provide rigorous safety guarantees due to the black-box nature of RL. 
Specifically, GLOSA requires two types of safety guarantees: (i) car-following safety, meaning that a CAV should not collide with its preceding vehicle, and (ii) red-light safety, meaning that a CAV should not run a red light. Existing RL-based GLOSA strategies can hardly provide theoretical guarantees for these two types of safety considerations. 
Although Ref. \cite{ding2024eco} explored the use of quadratic programming (QP) to convert RL actions to safe actions, the QP formulation only considers safety constraints at the current time step and thus may not ensure safety over time. 
Ref. \cite{waqas2022correct} proposed a less myopic approach, whereby a piecewise continuously differentiable time-varying CBF was devised to ensure red-light safety even at future steps. However, this work did not consider eco-driving nor integrate the proposed CBF with RL. 

\vspace{0.2em} \noindent \textbf{Statement of Contribution}. This paper develops a platoon-centric, safe RL-based GLOSA system to optimize the speed profile of CAVs in a mixed-traffic platoon of CAVs and HDVs. The contributions are two-fold. First, unlike existing vehicle-centric GLOSA systems that focus on individual CAVs, we jointly optimize the energy consumption of the entire platoon by controlling CAVs using multi-agent reinforcement learning (MARL). 
Second, unlike existing RL-based GLOSA policies that treat safety as a soft constraint in the reward function, we provide theoretical safety guarantees by integrating RL with Control Barrier Functions (CBFs) that explicitly characterize car-following safety and red-light safety. These CBFs are incorporated into the RL agent via a differentiable Quadratic Programming (QP) layer to convert the RL actions to safe actions, whose parameters can be jointly trained with the RL agent. The integration with CBFs ensures the safety of the mixed-traffic platoon over time. 

The rest of this paper is organized as follows. Section \ref{sec:Preliminary} provides preliminaries regarding MARL and CBF. Section \ref{sec:Prob} introduces the GLOSA problem. Section \ref{sec:method} describes our proposed safe RL-based method to control the vehicle platoon. Section \ref{sec:result} conducts simulations to illustrate the benefits of our proposed method compared to existing benchmarks. Section \ref{sec:conclusion} concludes the paper and presents future research directions.

\section{Preliminary}\label{sec:Preliminary}

In this section, we introduce multi-agent reinforcement learning in Section \ref{sec:Preliminary-RL} and control barrier function in Section \ref{sec:Preliminary-CBF} as theoretical foundations for the proposed approach.

\subsection{Multi-Agent Reinforcement Learning (MARL)}\label{sec:Preliminary-RL}
Multi-Agent Reinforcement Learning (RL) addresses the problem of learning to control a dynamic system based on current system observations via trial and error. In our work, the control problem is modeled as a decentralized partially observable Markov Decision Process (DEC-POMDP) executed by $n$ agents, denoted by $ \mathcal{M} = (\mathcal{S}, \mathcal{O},\mathcal{A}, \mathcal{P}, \{\bm{r}_i \}_{i=1}^n, \gamma) $. The state space $\mathcal{S}$ represents the set of all possible states that describe the entire environment, i.e., $\bm{s}_t \in \mathcal{S}$. Since all agents interact within one single environment, they share the same state space. The joint action space $\mathcal{A}$ represents the set of all possible joint actions taken by agents, i.e., $\bm{a}_t:= (\bm{a}_{0,t}, \bm{a}_{1,t}, \dots, \bm{a}_{n,t}) \in \mathcal{A}$. The observation space $\mathcal{O}$ comprises the union of the observation spaces of all agents, i.e., $\mathcal{O} = \bigcup\nolimits_i \mathcal{O}_i$, where $\mathcal{O}_i$ denotes the observation space of agent $i$. Each agent has access to only a partial representation of the global state $\bm{s}_t$ rather than the full state information. Formally, the observation of agent $i$ at time step $t$ is given by $\bm{o}_{i,t} = \mathcal{F} (\bm{s}_t; i) \in \mathcal{O}_i$, where $\mathcal{F}$ is an observation function that maps the global state $\bm{s}_t$ to the local observation $\bm{o}_{i,t}$. $\mathcal{P} $ represents the system transition probability from the current state $\bm{s}_t$ to the next state $\bm{s}_{t+1}$,  given the joint action $\bm{a}_t $. The reward function $\bm{r}_{i,t}: \mathcal{S} \times \mathcal{A} \rightarrow \mathbb{R}$ defines the reward of performing the joint action in the current state for agent $i$, and $\gamma \in (0,1] $ refers to the discount factor for future reward. 

Each agent has its own policy $\pi_{\theta_i}(\bm{a}_{i,t}|\bm{o}_{i,t})$ parametrized by $\theta_i$, which is defined as a probabilistic mapping from the current local observation to the current action. The goal of MARL is to derive these policy parameters $\{\theta_{i}\}_{i=1}^n$ that maximize the objective function $J(\pi) = \mathbb{E}_{\zeta \sim p_{\pi}}\Big[\sum^{T}_{t=0} \gamma^{t}r(\bm{s}_t, \bm{a}_t)\Big]$, where $ \zeta = (\bm{s}_0, \bm{o}_0, \bm{a}_0, ..., \bm{s}_T, \bm{o}_T, \bm{a}_T)$ denotes the sequence of states, observations, and joint actions; $p_{\pi}$ denotes the distribution of trajectories under the joint policy $\{\pi_{\theta_i}\}_{i=1}^n$.

In this paper, we leverage Multi-Agent Proximal Policy Optimization (MAPPO) to optimize these policies for CAVs \cite{yu2022surprising}. 
Each MAPPO agent $i$ incorporates an actor network parameterized by $\theta_i$ to generate the policy parameter and a critic network parameterized by $\varphi_i$ to estimate the state value $V_i(\bm{s}_t) $. Note that the critic network shares the same input and reward structure, consistent with the design of Centralized Training and Decentralized Execution (CTDE). This CTDE design leverages global information for efficient policy updates while enabling decentralized execution in partially observable and dynamic environments, such as mixed traffic, where CAVs cannot fully observe the entire environment state. Nevertheless, the reward values are distinct due to different emphases on the global reward. The actor updates its parameter according to the clipped policy loss function in Eq.(\ref{eq:actor-obj}).
\begin{equation} \label{eq:actor-obj}
    \mathcal{L}^{\text{clip}}(\theta_i) = \hat{\mathbb{E}}_{t}\Big[\min \big(r_{i,t}(\theta_i) \hat{A}_{i,t}, \text{clip}(r_{i,t}(\theta_i), 1-\epsilon, 1+\epsilon)\hat{A}_{i,t} \big) \Big],
\end{equation}
where $r_{i,t}(\theta_i)=\frac{\pi_{\theta_i}(\bm{a}_{i,t}| \bm{o}_{i,t})}{\pi_{\theta_{i,old}}(\bm{a}_{i,t}| \bm{o}_{i,t})}$ represents the probability ratio; $\hat{A}_{i,t}= \bm{r}_{i,t} + \gamma \hat{V}_{\varphi_i}(s_{t+1}) - \hat{V}_{\varphi_i}(s_{t})$ represents the the advantage term, and $ \hat{V}_{\varphi_i}(\bm{s}_t) $ is the estimated state value given by the critic. The critic updates its parameter to provide a more accurate estimation for the state value function $V_i(\bm{s}_t)$ as: 
\begin{equation} \label{eq:critic-obj}
    \mathcal{L}(\varphi) =
\begin{cases}
0.5 \delta ^ 2 & |\delta| < 1 \\
|\delta| - 0.5 & |\delta| \ge 1
\end{cases},
\end{equation}
where we have $\delta:= \bm{r}_{i,t} + \gamma \hat{V}_{\varphi_i}(s_{t+1}) - \hat{V}_{\varphi_i}(s_{t})$.

\subsection{Control Barrier Function}\label{sec:Preliminary-CBF}

Control Barrier Functions (CBFs) are used to ensure that the system state remains within a predefined safety set by imposing constraints on the control inputs. Specifically, let us consider a control affine system defined by Eq.(\ref{eq:affine-sys}). 
\begin{equation}\label{eq:affine-sys}
    \dot{\bm{s}} = f(\bm{s}) + g(\bm{s})\bm{a},
\end{equation}
where $ \bm{s} \in \mathcal{S} \subset \mathbb{R}^{n}$ and $ \bm{a} \in \mathcal{A} \subset \mathbb{R}^{m}$ are the system state and control input, respectively; $ f(\bm{s}): \mathbb{R}^{n} \rightarrow \mathbb{R}^{n}$ and $ g(\bm{s}): \mathbb{R}^{n} \rightarrow \mathbb{R}^{n \times m}$ represents the system dynamics and input dynamics, respectively. Here, $f(\bm{s})$ and $g(\bm{s})$ are locally Lipschitz continuous. CBF  for this control affine system is defined by Definition \ref{def:CBF}.

\begin{Definition}[Control Barrier Function \cite{ames2016control}] \label{def:CBF}
A continuously differentiable function $ h(\bm{s}):\mathcal{S} \rightarrow \mathbb{R}$ is a control barrier function for a control affine system if the system state is required to be maintained in a forward-invariant safety set $\mathcal{C}:= \big\{\bm{s} \in \mathcal{S} \subset \mathbb{R}^{n}\big| h(\bm{s}) \ge 0 \}$ characterized by $h(\bm{s})$.
\end{Definition}

The barrier condition of CBF $h(\bm{s})$ is written as Eq.(\ref{eq:Barrier Condition}).
\begin{equation} \label{eq:Barrier Condition}
\sup_{a \in \mathcal{A}} \big[ L_f h(\bm{s}) + L_g h(\bm{s}) \bm{a}  \big] \geq - \alpha(h(\bm{s})),
\end{equation}
where $L_f h(\bm{s}) $ and $ L_g h(\bm{s})$ represents the Lie derivatives of the function $h(\bm{s})$ with respect to the system dynamics $f(\bm{s})$ and $g(\bm{s})$, i.e., $L_f h(\bm{s}):= \big(\frac{\partial h(\bm{s})}{\partial \bm{s}}\big)^{\mathsf{T}} f(\bm{s}) $ and $L_g h(\bm{s}):= \big(\frac{\partial h(\bm{s})}{\partial \bm{s}}\big)^{\mathsf{T}} g(\bm{s}) $; and $\alpha$ is an extended class $\mathcal{K}_{\infty}$ functions\footnote{a continuous function $\alpha$ belongs to an extended class
$\mathcal{K}_{\infty}$ if it is strictly increasing and $\alpha(0) = 0$}. Note that this barrier condition can be equivalently transformed as  
\begin{align}
   \sup_{a \in \mathcal{A}} \frac{dh}{dt} &= \sup_{a \in \mathcal{A}} \big [\big(\frac{\partial h(\bm{s})}{\partial \bm{s}}\big)^{\mathsf{T}} \big(f(\bm{s}) + g(\bm{s})\bm{a}\big)\big] \notag \\
    &=  \sup_{a \in \mathcal{A}} \big [ L_f h(\bm{s}) + L_g h(\bm{s}) \bm{a} \big] \geq - \alpha(h(\bm{s})).   \label{eq:Barrier Condition-time}
\end{align}
By Eq.\eqref{eq:Barrier Condition-time}, the barrier condition ensures that if the current state remains in set $\mathcal{C}$, with appropriate control input $\bm{a}$, the subsequent states are guaranteed to stay within the safety set~$\mathcal{C}$.

\section{Problem Statement} \label{sec:Prob}

\begin{figure*}[h]
    \centering
    \includegraphics[width= 0.6\textwidth]{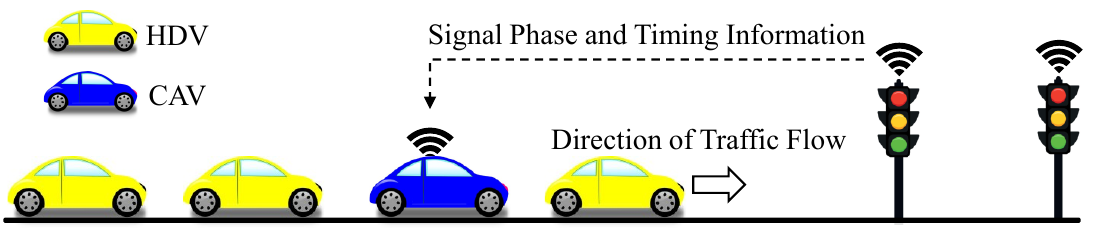}
    \caption{Problem setup. A mixed-traffic platoon composed of CAVs and HDVs travels along an arterial road with multiple pre-timed intersections. Each CAV makes decisions based on information from its immediate preceding and following vehicles, obtained through onboard sensors, as well as signal phase and timing information acquired through vehicle-to-infrastructure communication.
}
    \label{fig:scenario}
\end{figure*}

We formally formulate the GLOSA problem for mixed-traffic platoons. Let us discretize the considered time horizon as a set of time steps denoted by $ \mathcal{T}:= \{0, \Delta T, 2\Delta T, ..., T\}$. 
As shown in Figure \ref{fig:scenario}, we consider a mixed-traffic platoon, including both CAVs and HDVs, that intends to pass an arterial consisting of $K$ signalized intersections denoted by $\mathcal{K}:= \{0, 1, ..., K-1\}$. Each intersection is implemented with a fixed-time signal control policy. 
Let $p_k$ denote the position of the $k^{th}$ traffic signal light the platoon encounters. The timing sequence of signal $k$ is represented as $\{g_{k0}, r_{k0}, g_{k1}, r_{k1}, g_{k2}, r_{k2}, ...\}$, where $g_{kj}$ and $r_{kj}$ denote the $j^{th}$ transition of the signal $k$ to green and red, respectively, which are assumed to be integral multiples of $\Delta T$. Here, we only consider red and green signal phases for simplicity. Other common signal indications, e.g., yellow, are considered red. 
 
Let $\mathcal{N}:= \{0, 1,..., N\}$ denote the vehicles in the platoon. For each vehicle $i\in \mathcal{N}$, we denote its position and speed at time step $t$ by $x_{i}(t)$ and $v_{i}(t)$, respectively. We make the following remarks about these vehicles. First, we consider a scenario with low connectivity in which CAVs can only receive information about real-time signal timings from the infrastructure via I2V communications, and HDVs are non-connected and unable to provide position and speed information. Such a low-connectivity scenario is practical as the installation of V2X infrastructure can be costly and will take time. Consequently, each CAV may only obtain the position and speed information of its immediate preceding and following vehicles from its onboard sensors \cite{sarker2019review}. 
Second, we assume that the automation level of CAVs is sufficient to enable the automatic execution of the received longitudinal action, e.g., SAE Level 2. This assumption is realistic, as such vehicles are becoming mainstream in the market \cite{sekadakis2025systematic, ding2024exploratory}.
Third, all CAVs are assumed to be equipped with our proposed controller, and HDVs will adjust their speed according to the state of the preceding vehicle (e.g., headway and speed) based on their car-following behavior. Here, we use the commonly used car-following model, i.e., the Intelligent Driver Model (IDM), to simulate the driving behavior of HDVs. The main reason for using this IDM model is that it can realistically simulate real-world human driving behavior \cite{he2022coordinated, xu2024multi}. 
Last, we do not consider lane changes in our scenario for simplicity and to ensure stable RL training. However, our proposed method can be extended to consider lane changes by adding or removing vehicles in the vehicle set $\mathcal{N}$ \cite{liu2024multistep, wu2024continuous}.

Next, we model the operation of the mixed-traffic platoon. The evolution of the states of each vehicle can be updated by
\begin{equation}\label{eq:sys-dynamics}
\begin{cases}
\dot{x}_i(t) = v_i(t), \\
\dot{v}_i(t) = a_i(t),
\end{cases}
\end{equation}
where the acceleration of a CAV is determined by our safe RL-based controller, while the acceleration of each HDV is given by an IDM model defined by Eq.(\ref{eq:IDM}). 
\begin{equation}\label{eq:IDM}
    a_i(t) = a_0\left(1 - \left(\frac{v_i(t)}{v^{\rm{max}}}\right)^{\delta_0} - \left(\frac{s^*(v_i(t), \Delta v_i(t))}{s_i(t)}\right)^2\right), 
\end{equation}
with 
\begin{equation}
    s^*(v_i(t), \Delta v_i(t)):= s_{0} + \max \left(0, Tv_i (t) + \frac{ v_i(t) \Delta v_i(t)}{2\sqrt{a_0 b_0}}\right),
\end{equation}
where $v^{\rm{max}}$ is the maximum allowed speed; $\delta_0$ is the acceleration exponent; $s_i(t)$ and $s^*$, respectively, denote the actual and desired gap distances; $s_0$ represents the minimum gap distance; $T$ is the desired minimum headway to the preceding vehicle; $\Delta v_i$ is the speed
difference to the preceding vehicle; and $a_0$ and $b_0$ denote the maximum acceleration and deceleration for HDVs, respectively. 
To ensure no HDVs run red lights, we consider HDVs to follow a phantom vehicle stopped in front of the stop line during red time. 

\begin{figure}[h]
    \centering
	\includegraphics[width=0.8 \columnwidth]{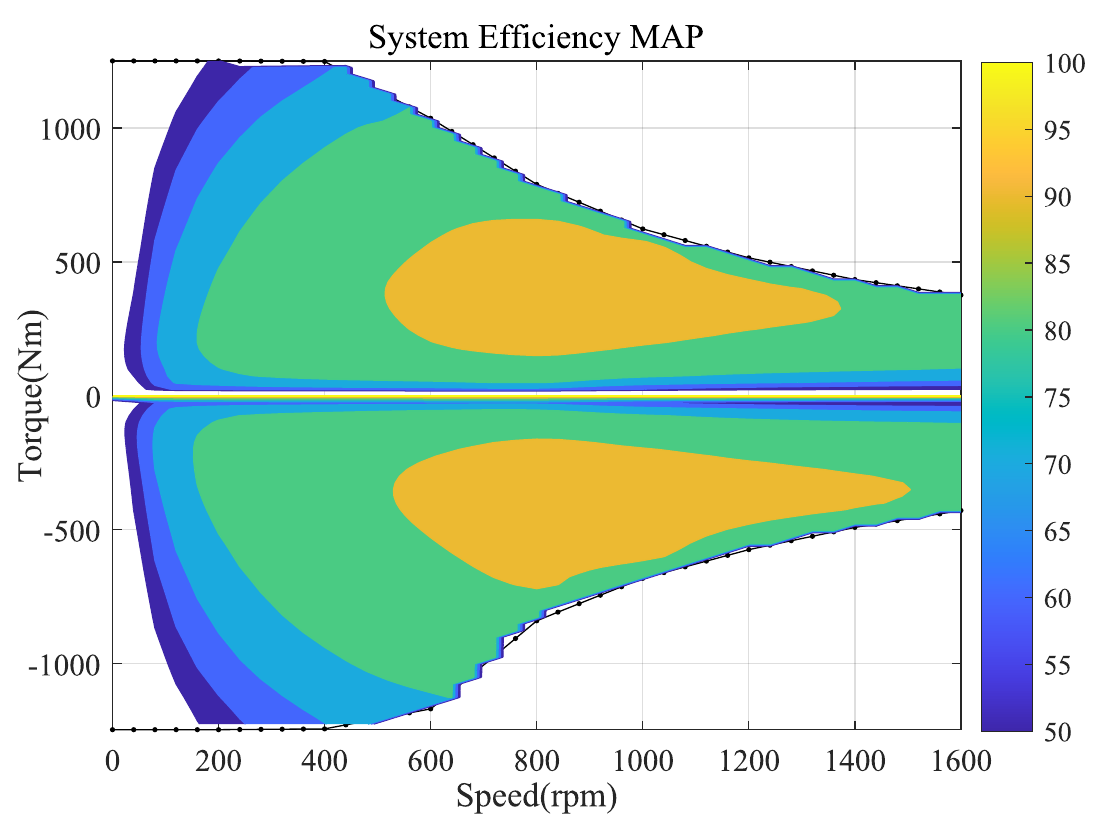}
    \caption{Motor efficiency map. Motor efficiency in EVs depends on speed and torque, through which energy consumption can be derived.}
    \label{fig: motor-efficiency-map}
\end{figure}

In this paper, we consider all vehicles to be Electric Vehicles (EVs), following the trend of sustainable transportation systems \cite{khalatbarisoltani2024safety}. The energy consumption is computed according to Eq.(\ref{eq:energy-ev}).
\begin{equation}\label{eq:energy-ev}
   EC_{i}(t) = F_{i}(t)v_i(t)\Delta T \eta^{-\sigma(F_i(t))} \sigma(F_i(t)),
\end{equation}
with driving force $ F_i(t) $ computed by
\begin{equation}\label{eq:power-ev}
   F_i(t)= ma_i(t) + f_a v_i^2(t) + mgf_r\cos \alpha + mg\sin\alpha,
\end{equation}
where $\sigma(\cdot)$ is the sign function; $m$ is the mass of electric vehicles; $g$ is the gravitational acceleration; $f_a$ and $f_r$ are drag coefficient and wheel rolling resilience coefficient, respectively; $\alpha$ represents road gradient; and $\eta $ is the energy conversion efficiency of driving and braking power. We assume that all EVs are powered by Protean PD18 RAM motors \cite{liang2023energy}. These motors have a well-characterized motor efficiency map as shown in Figure \ref{fig: motor-efficiency-map}, which explicitly reflects the nonlinear nature of motor efficiency. The energy conversion efficiency can thus be determined by the torque $ Tq:= F_i(t)r $ and vehicle speed $ v_i(t) $ from the motor efficiency map, where $ r $ is the radius of the vehicle tire. Since our energy consumption model explicitly captures the energy recycling process during braking, our proposed method can be generalized to a broad class of EV propulsion systems without loss of generality.

\section{Methodology}\label{sec:method}
\begin{figure*}[h]
    \centering
    \includegraphics[width= 0.7\textwidth]{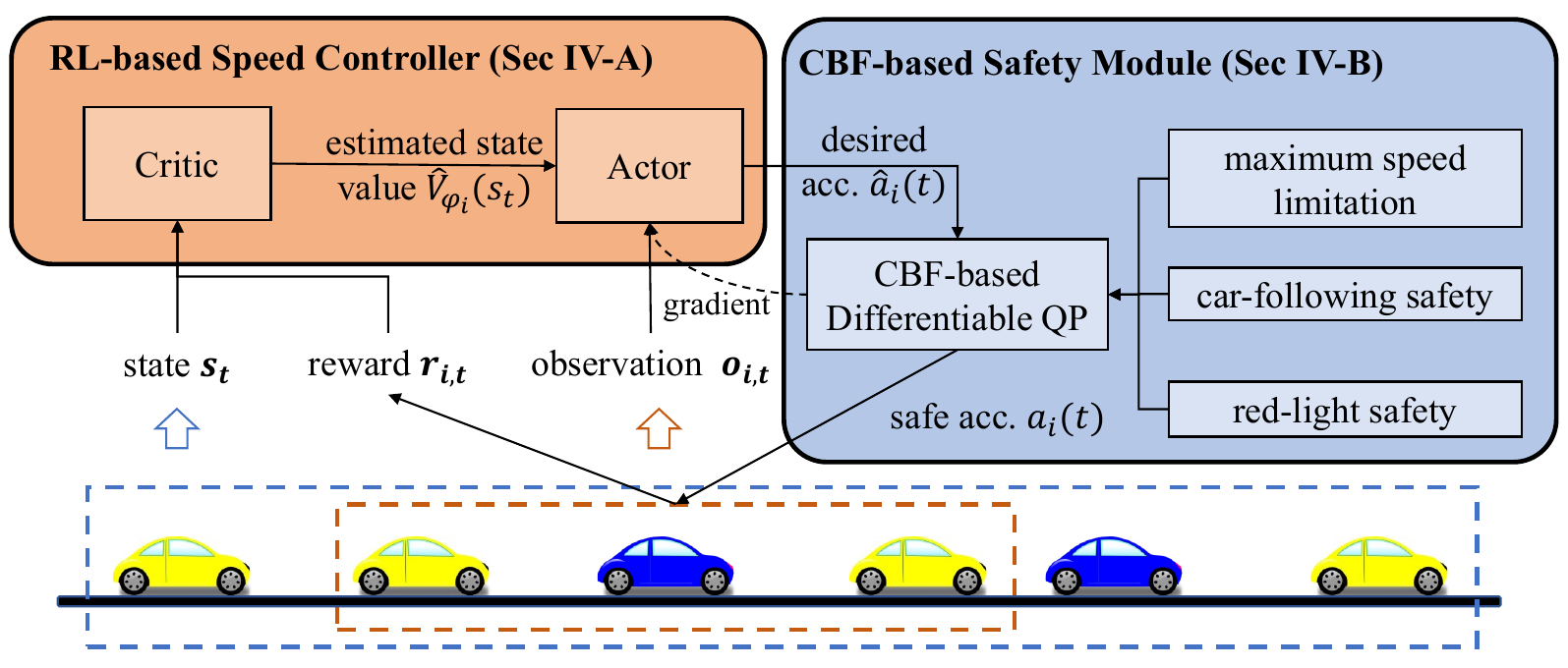}
    \caption{Overview of safe RL-based control framework. At each time step, the $i$-th RL agent receives the system observation $\bm{o}_{i,t}$ as input and outputs a desired acceleration $\hat{a}_i(t)$, abbreviated as \textit{desired acc.} in the figure. The CBF-based QP then refines this acceleration, generating a safe acceleration $a_i(t)$, abbreviated as \textit{safe acc.} in the figure. Meanwhile, the critic estimates the state value $\hat{V}_{\phi_i}(\bm{s}_t)$ based on the system state $\bm{s}_t$ for policy updates. After executing the safe action in the environment, the agent receives a reward signal $\bm{r}_{i,t}$, which is used to update the critic.
}
    \label{fig:algo}
\end{figure*}

This section introduces our proposed platoon-centric, safe RL-based GLOSA framework to control a mixed-traffic platoon of CAVs and HDVs. The methodological framework is illustrated in Figure \ref{fig:algo}, where each CAV is controlled by an RL agent, including (i) an RL-based speed controller (the red part) that generates desired acceleration and (ii) a CBF-based safety module (the blue part) that provides safety assurance. Compared to existing works that only focus on the ego CAV, the RL-based controller is platoon-centric in that it explicitly considers the impacts of the CAV on the energy consumption of the following vehicles, including other CAVs and unobservable HDVs. Since the RL-based controller serves as a black box, we devise the CBF-based safety module that uses CBFs to characterize the requirement of car-following safety and red-light safety, which are incorporated into a quadratic programming (QP) layer to convert RL actions to safe actions. 

Next, we present the details of the methodological framework. Section \ref{Sec:RL-based Controller} introduces our RL-based speed controller, Section \ref{Sec:Trajectory Planning Model} formulates the car-following safety and red-light safety conditions, and Section \ref{Sec:Differentiable QP} illustrates differentiable QP that utilizes the gradient information to update the RL agent.

\subsection{RL-based Speed Controller}\label{Sec:RL-based Controller}
The CAV control problem is formulated as a DEC-POMDP $ \mathcal{M} = (\mathcal{S}, \mathcal{O}, \mathcal{A}, \mathcal{P}, \{\bm{r_i}\}_{i=1}^n, \gamma) $, where each CAV is controlled by an RL agent. The RL agent is trained using MAPPO with an actor-critic framework, whereby an actor network $\pi_{\theta_i}(\bm{a}_t\mid \bm{s}_t)$ generates the action, which is evaluated by a critic network $V_{\phi_i}(\bm{s}_t) $.  
For each CAV, the elements are introduced as follows:  

\textbf{State $\bm{s}_t \in \mathcal{S}$}. The global state provides a representation of the entire traffic environment that involves all vehicles and signalized intersections. Specifically, the state includes (i) the speed and spacing of both HDVs and CAVs, (ii) the distance of CAVs to the next traffic light, and (iii) the signal indications and the remaining duration of the current phase of all traffic lights. The state is used as input to the critic network during centralized training to enhance stability and accelerate convergence.

\textbf{Observation $\bm{o}_{i,t} \in \mathcal{O}_i \subset \mathcal{O}$}. As CAVs can only perceive a limited portion of the environment, they make decisions based solely on their own observations, which serve as inputs to the actor networks. The observation of a specific CAV consists of (i) the spacing of this CAV to their immediate preceding and following vehicles, as well as the speed of these three vehicles, and (ii) the distance to the closest traffic light, along with the signal indication and remaining phase duration of this traffic light.

\textbf{Action $\bm{a}_{i,t} \in \mathcal{A}$}. We define the action of a specific CAV $i \in \mathcal{N}$ by the desired acceleration $\hat{a}_i(t)$ that the CAVs aim to achieve, sampled from policy $\pi_{\theta_i}(\bm{a}_{i,t}| \bm{o}_{i,t})$.

\textbf{Reward $\bm{r_i}$}. At each time step, the agent receives a reward from the environment. The reward for a specific vehicle $i$ is written as Eq.(\ref{eq:step-reward}).
\begin{equation}\label{eq:step-reward}
    \bm{r}_{i, t} = w_1 r^{\rm{safety}}_{i,t} + w_2 r^{\rm{efficiency}}_{i,t} + w_3 r^{\rm{stability}}_{i,t} + w_4 r^{\rm{energy}}_{i,t},
\end{equation}
where $w_1$, $w_2$, $w_3$, and $w_4$ are the weights of the safety reward, efficiency reward, platoon stability reward, and energy consumption reward, respectively. These weights are chosen via a trial-and-error approach to achieve desired properties. First, a larger weight is assigned to the travel efficiency reward to encourage the forward movement of CAVs \cite{zhou2019development}. While safety and stability remain the primary focus of our study, excessively high weights for these terms could lead to undesired vehicle stoppage. Second, the weight of the energy reward is set lower since the energy consumption of following vehicles amplifies the magnitude of the energy reward \cite{liang2023energy}.

The safety reward aims to encourage the desired acceleration $\hat{a}_i(t)$ to fall within a safety region. In case the desired acceleration is not safe, we generate a safe acceleration $a_i(t)$ closest to $\hat{a}_i(t)$ that satisfies safety conditions defined by CBF-based QP introduced in Section \ref{Sec:Trajectory Planning Model}. Therefore, we define the safety reward by the difference between the desired acceleration $\hat{a}_{i}(t)$ and the safe acceleration $a_{i}(t)$ as Eq.(\ref{eq:safe-reward}).
\begin{equation} \label{eq:safe-reward}
    r^{\rm{safety}}_{i,t} = - \left(\hat{a}_{i}(t) - a_{i}(t)\right)^2.
\end{equation}

The efficiency reward is defined as the vehicle speed, i.e., 
\begin{equation}\label{eq:efficiency-reward}
   r^{\rm{efficiency}}_{i,t} = v_i(t).
\end{equation}
This is because the speed can reflect the distance traveled by each vehicle per unit of time, which encourages the vehicles to travel faster and pass the arterial within the given time horizon.

The platoon stability reward aims to prevent frequent acceleration and deceleration. Instead of only considering the CAV itself, we sum up the weighted square of accelerations of all its following vehicles as Eq.(\ref{eq:stability-reward}).
\begin{equation}\label{eq:stability-reward}
    r_{i,t}^{\rm{stability}} = - \sum_{j \in \mathcal{N}_{i}^{\rm{F}}/\{i\}} \kappa^{i-j} a_j^2(t),
\end{equation}
where the impact factor $\kappa$ represents the influence of CAV $i$ on its following vehicles, and $\mathcal{N}_{i}^{\rm{F}}$ is the set of all vehicles following vehicle $i$. 

The energy consumption reward is defined similarly from a platoon perspective. Specifically, we compute the energy consumption reward as Eq.(\ref{eq:energy-reward}).
\begin{equation}\label{eq:energy-reward}
    r_{i,t}^{\rm{energy}} = - \sum_{j \in \mathcal{N}_{i}^{\rm{F}}} \kappa^{i-j} EC_j(t).
\end{equation}

The RL-based speed controller is trained using an actor-critic algorithm with privileged information to handle the uncertainties brought by unobserved HDVs. Specifically, during training, the critic uses the full state $\bm{s}_t$ (rather than $\bm{o}_{i,t}$) to estimate the state value, which is not available during deployment. This privileged information helps the critic learn more effectively by providing a more comprehensive understanding of the environment, thereby improving the robustness of the RL-based controller when it is deployed without access to this privileged information.

\subsection{CBF-Based QP for Car-following and Red-Light Safety}\label{Sec:Trajectory Planning Model}

To enhance safety, we convert the RL action $\hat{a}_i(t)$ to a safe action via a CBF-based QP. Specifically, we seek to obtain an acceleration $a_i(t)$ that minimizes the deviation between these accelerations, i.e., $\left(\hat{a}_i(t)-a_i(t)\right)^2$, while satisfying the following constraints: (i) physical limitations, e.g., maximum allowable acceleration/deceleration, maximum speed, etc., (ii) car-following safety, and (iii) red-light safety. Here, the car-following and red-light safety constraints are characterized by CBFs, which will be detailed below. 

\subsubsection{Car-following safety}
For car-following safety, we design a CBF candidate as presented in Eq.(\ref{eq:cf-safety}). 
Here, we adopt the concept of maintaining a desired gap distance as defined in the IDM model. Specifically, Eq.(\ref{eq:cf-safety}) ensures that the difference between the desired and actual gap distances remains positive.
\begin{equation}\label{eq:cf-safety}
    h_1(t) = x_{i-1}(t) - x_{i}(t) - \tau v_i(t) - L - S_0 - S^{\rm{des}}(t),
\end{equation}
where $L$ is the vehicle length; $\tau$ represents the desired time headway; $ S_0 $ is the safe stopping distance; $S^{\rm{dec}}(t)$ is the deceleration distance calculated by Eq.(\ref{eq:deceleration-dist}).
\begin{equation}\label{eq:deceleration-dist}
    S^{\rm{dec}}(t) = 
    \left\{
    \begin{array}{ll}
        \dfrac{(v_{i-1}(t) - v_i(t))^2}{2a^{\rm{min}}}, & \text{if } v_i(t) \ge v_{i-1}(t) \\
        0, & \text{otherwise}
    \end{array}
    \right.
\end{equation}
where $a^{\rm{min}}$ is the negative value of the maximum deceleration of CAVs.

Let us define $v_i^{\rm{r}}(t):= v_{i-1}(t)-v_i(t)$ as the relative speed between vehicle $i-1$ and $i$. Then, by Eq.\eqref{eq:Barrier Condition-time}, the barrier condition of $h_1(t)$, i.e., $\frac{dh_1(t)}{dt}\geq -\alpha(h_1(t))$, can be written as Eq.\eqref{eq:CBF-h1} by explicitly calculating the derivatives.  
\begin{equation}\label{eq:CBF-h1}
    \left\{
    \begin{array}{ll}
        \left(\tau - \dfrac{v_{i}^{\rm r}(t)}{a^{\rm min}}\right) a_i(t) \le h_1(t) + v_{i}^{\rm r}(t), & \text{if } v_i(t) \ge v_{i-1}(t) \\
        \tau a_i(t) \le h_1(t) + v_{i}^{\rm r}(t), & \text{otherwise}
    \end{array}
    \right.
\end{equation}
Here, we specify the $\mathcal{K}_{\infty}$ class function to be linear with a coefficient of $1$ following the insights of Ref. \cite{waqas2022correct}, which can achieve a balance between safety and efficiency.

\subsubsection{Red-light safety}
For red-light safety, we seek to ensure that CAVs will not run red lights. The key idea is to add constraints to the passing time of each CAV such that it always passes through the intersection during green indications. To this end, we consider two possible scenarios depending on the signal phase in which the passing time is located.  

\begin{figure}[h]
    \centering
    \includegraphics[width= 0.8 \columnwidth]{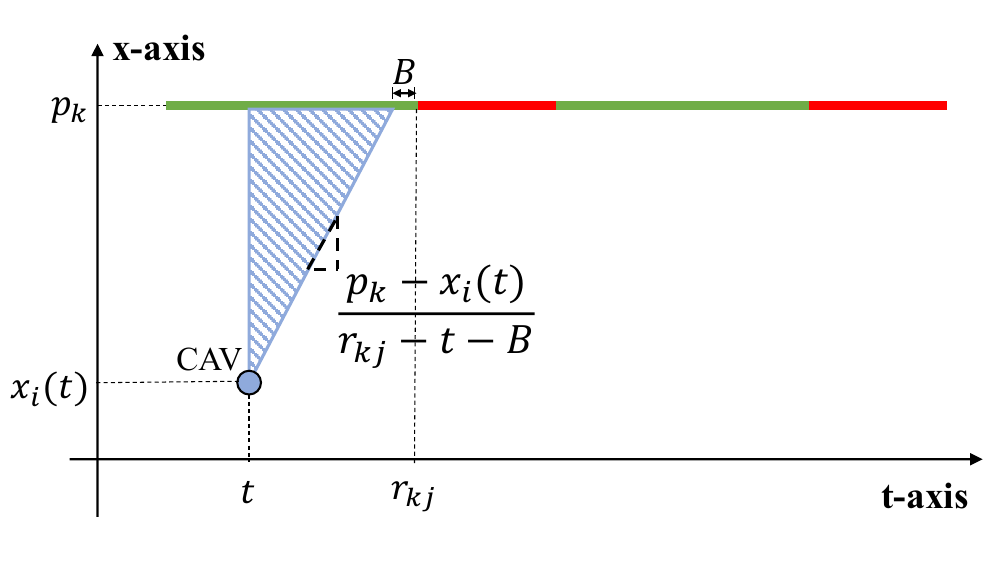}
    \caption{Passing scenario 1 with a feasible speed region (blue part). The CAV approaches the intersection during the green phase and successfully passes through it within the same phase.}
    \label{fig:tl_case1}
\end{figure}

\noindent \textbf{Scenario 1}. As shown in Figure \ref{fig:tl_case1}, in this scenario, the CAV (indexed by $i$) approaches the $k^{th}$ intersection within a green interval and will pass within the current green phase. The key requirement in this scenario is to have a buffer $B$ between the passing time of the CAV and the start time of the red indication so that the CAV is guaranteed not to run a red light. 

Consequently, we design the CBF candidate as Eq.(\ref{eq:tl-case1pt}), which ensures that 
the remaining phase duration is longer than the desired headway when the CAV passes the intersection. 
\begin{equation}\label{eq:tl-case1pt}
  h_2^{\rm{tl1}}(t) = v_i(t) - \frac{p_k-x_{i}(t)}{r_{kj} - t - B}.
\end{equation}
 The barrier condition of Eq.(\ref{eq:tl-case1pt}), i.e., $\frac{dh_2^{\rm{tl1}}(t)}{dt}\geq -\alpha(h_2^{\rm{tl1}}(t))$, can be rewritten as Eq.(\ref{eq:CBF-h2-case1}) by calculating the time derivatives of $h_2^{\rm{tl1}}$.

\begin{align}
    a_i(t) \ge \frac{p_k-x_{i}(t)-v_i(t)(r_{kj} - t - B)}{(r_{kj} - t - B)^2} - h_2^{\rm{tl1}}(t).\label{eq:CBF-h2-case1}
\end{align}

\begin{figure}[h]
    \centering
    \includegraphics[width= 0.8 \columnwidth]{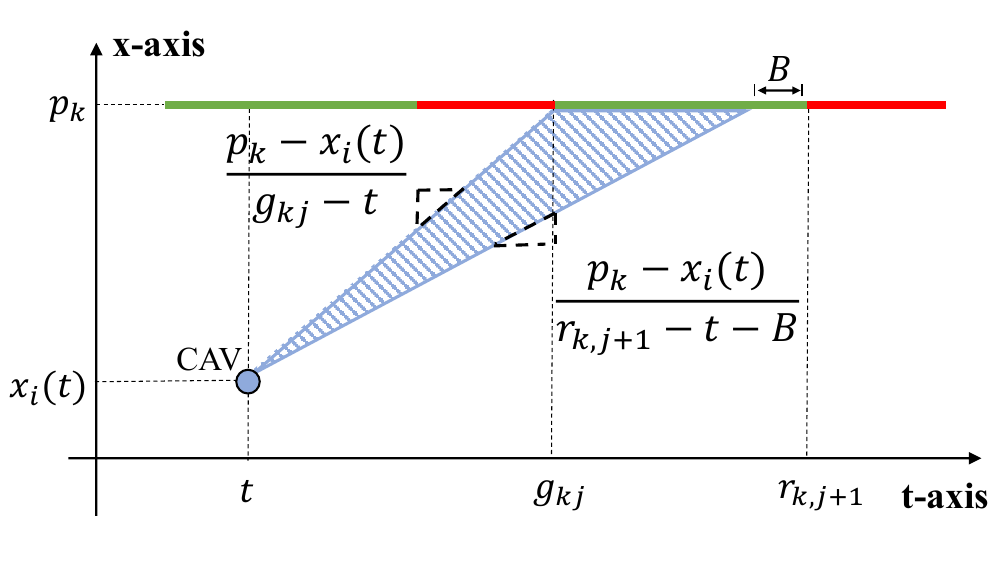}
    \caption{Passing scenario 2 with a feasible speed region (blue part). The CAV approaches the intersection during either the green or red phase but will pass through it in the next green phase.}
    \label{fig:tl_case2}
\end{figure}

\noindent \textbf{Scenario 2.} As shown in Figure \ref{fig:tl_case2}, in this scenario, the CAV is expected to pass the $k^{th}$ intersection within the next green phase, regardless of the current traffic light indication. Similar to Scenario 1, we construct CBFs to ensure that the CAV passes through the intersection between the green start time $g_{kj}$ and the red start time $r_{k,j+1}$ with a buffer $B$. 

Specifically, we derive the CBF candidate to provide an upper bound to the speed of the CAV, which ensures that the CAV passes through the intersection after $g_{kj}$.
\begin{equation}\label{eq:tl-case21}
  h_{2.1}^{\rm{tl2}}(t) = \frac{p_k-x_i(t)}{g_{kj}-t} - v_i(t). 
\end{equation}
By Eq.(\ref{eq:Barrier Condition-time}), the barrier condition of Eq.(\ref{eq:tl-case21}), i.e., $\frac{dh_{2.1}^{\rm{tl2}}(t)}{dt} \geq - \alpha(h_{2.1}^{\rm{tl2}}(t))$, is written as Eq.(\ref{eq:CBF-h2-case21}) by expanding the derivatives.

\begin{align}
    a_i(t) \le \frac{p_k-x_{i}(t)-v_i(t)(g_{kj} - t)}{(g_{kj} - t)^2} + h_{2.1}^{\rm{tl2}}(t).\label{eq:CBF-h2-case21}
\end{align}

Similarly, we derive the CBF candidate to provide a lower bound to the speed of the CAV, which ensures that the CAV passes through the intersection before $r_{k,j+1} - B$. 
 \begin{equation}\label{eq:tl-case22}
h_{2.2}^{\rm{tl2}}(t) = v_i(t) - \frac{p_k-x_i(t)}{r_{k,j+1} - t - B}.
\end{equation}
The barrier condition of Eq.(\ref{eq:tl-case22}) is written as Eq.(\ref{eq:CBF-h2-case22}).
\begin{align}
    a_i(t) \ge \frac{p_k-x_{i}(t)-v_i(t)(r_{k,j+1} - t - B)}{(r_{k,j+1} - t - B)^2} - h_{2.2}^{\rm{tl2}}(t).\label{eq:CBF-h2-case22}
\end{align}

Note that Scenario 1 and Scenario 2 are common scenarios where vehicles will pass the intersection only during the current green phase or the next green phase. Our method is ready to be extended to the case where vehicles pass the intersection during subsequent green intervals, which we ignore in this paper for simplicity.

To sum up, we can construct a 
CBF-based QP problem as Eq.(\ref{eq:CBF-QP-obj})-Eq.(\ref{eq:CBF-QP-neg}), which converts the RL actions to safe actions using the CBF-based safety constraints described above. 
\begin{subequations}
    \begin{align}
    \min \limits_{\{a_{i}(t)\}} \quad & \big(a_{i}(t) - \hat{a}_{i}(t) \big)^2\label{eq:CBF-QP-obj}\\
    \mbox{s.t.} \quad & a_i(t) \Delta T \le v^{\rm{max}} - v_i(t) \label{eq:CBF-speed-limit}\\
    & \text{Eq.(\ref{eq:CBF-h1})\label{eq:CBF-QP-cfsafe}}\\
    & \text{Eqs.(\ref{eq:tl-case1pt})--(\ref{eq:CBF-h2-case1})~or~}
        \text{Eqs.(\ref{eq:tl-case21})--(\ref{eq:CBF-h2-case22})} \label{eq:CBF-QP-tlsafe}\\
    & -a^{\rm{min}}  \le a_i(t) \le a^{\rm{max}} \label{eq:CBF-QP-neg}
    \end{align}
\end{subequations}
where the constraint Eq.(\ref{eq:CBF-speed-limit}) indicates that after accelerating, the speed of CAVs should not exceed the speed limit. The constraint Eq.(\ref{eq:CBF-QP-cfsafe}) ensures car-following safety by taking into account the spacing between CAVs and preceding vehicles. Constraint Eq.(\ref{eq:CBF-QP-tlsafe}) ensures that vehicles should not run red lights by allowing them to reach the intersection during the green interval. Constraint Eq.(\ref{eq:CBF-QP-neg}) is the maximum acceleration and deceleration constraint. Note that the constraint Eq.(\ref{eq:CBF-QP-tlsafe}) has an \emph{or} operator, which in general introduces a binary variable. Nevertheless, we address this issue by solving one QP with Eq.(\ref{eq:CBF-h2-case1}) and the other QP with Eq.(\ref{eq:CBF-h2-case21}) and Eq.(\ref{eq:CBF-h2-case22}), and then choosing the solution that yields a lower optimum. Note that such a QP is easy to solve using classical interior-point algorithms because of its inherent convexity and low dimensionality.

\subsection{Differentiable QP Layer}\label{Sec:Differentiable QP}

Next, we introduce a differentiable QP layer that allows the backpropagation of the CBF-based QP solution. Such a treatment can facilitate training by allowing a better integration between the CBF-based QP and the RL framework. 

Let $u_i(t):= a_{i}(t) - \hat{a}_{i}(t)$ be the difference between the safe acceleration and the desired acceleration. Then, with a CBF-based QP, the partial derivative of actor loss can be obtained based on the chain rule as Eq.(\ref{eq:differentiable-QP-loss}).
\begin{align}
    \frac{\partial \ell}{\partial \theta_i} & =\frac{\partial \ell}{\partial a_{i}(t)} \frac{\partial a_{i}(t)}{\partial \theta_i} = \frac{\partial \ell}{\partial a_{i}(t)} \left(\frac{\partial \hat{a}_{i}(t)}{\partial \theta_i} + \frac{\partial u_i(t)}{\partial \theta_i}\right) \notag \\
    & = \frac{\partial \ell}{\partial a_{i}(t)} \left(\frac{\partial \hat{a}_{i}(t)}{\partial \theta_i} + \frac{\partial u_i(t)}{\partial \hat{a}_{i}(t)}\frac{\partial \hat{a}_{i}(t)}{\partial \theta_i}\right), \label{eq:differentiable-QP-loss}
\end{align}
where the first and third equations stem from the chain rule, and the second equation is based on the definition of $u_i(t)$. Note that here $\frac{\partial \ell}{\partial a_{i}(t)}$ can be obtained from the loss function, and $\frac{\hat{a}_{i}(t)}{\partial \theta_i}$ is determined by the architecture of the actor-network. Therefore, we only need to calculate $\frac{\partial u_i(t)}{\partial \hat{a}_{i}(t)}$.

We can derive $\frac{\partial u_i(t)}{\partial \hat{a}_{i}(t)}$ by differentiating the KKT condition. Let us rewrite the aforementioned CBF-based QP in a matrix form with a decision variable $\bm{u}$ as Eq.(\ref{eq:differentiable-QP-obj})-Eq.(\ref{eq:differentiable-QP-cons}).
\begin{subequations}\label{eq:differentiable-QP-form}
    \begin{align}
    \min \limits_{\{\bm{u}\}} \quad & \frac{1}{2}\bm{u}^{\mathsf{T}} Q \bm{u}\label{eq:differentiable-QP-obj}\\
    \mbox{s.t.} \quad & G \bm{u} \le F\big(\hat{a}_{i}(t)\big)\label{eq:differentiable-QP-cons}
    \end{align}
\end{subequations}
where $Q$ and $G$ are the corresponding parameter matrices of the QP problem; only $F\big(\hat{a}_{i}(t)\big)$ is the function of desired acceleration $\hat{a}_{i}(t)$. The Lagrangian of Eq.(\ref{eq:differentiable-QP-form}) is given by:
\begin{equation}\label{eq:QP-Lagrangian}
L(\bm{u}, \bm{\lambda}) = \frac{1}{2}\bm{u}^{\mathsf{T}} Q \bm{u} + \bm{\lambda} \Big( G \bm{u} - F\big(\hat{a}_{i}(t)\big)\Big),
\end{equation}
where $\lambda$ is the dual variable. The optimality conditions for this Lagrangian function are given as Eq.(\ref{eq:Lagrangian-optimality}).
\begin{equation}\label{eq:Lagrangian-optimality}
  \begin{aligned}
    \frac{\partial L}{\partial \bm{u}}(\bm{u}^*, \bm{\lambda}^*) &= Q\bm{u}^* + G^{\mathsf{T}} \bm{\lambda}^* = 0, \\
    \frac{\partial L}{\partial \bm{\lambda}}(\bm{u}^*, \bm{\lambda}^*) &= \mathbb{D}(\bm{\lambda}^*) \left( G \bm{u}^* - F\big(\hat{a}_{i}(t)\big) \right) = 0
  \end{aligned}
\end{equation}
where $\mathbb{D} \left(\bm{\lambda}^*\right)$ creates a diagonal matrix from $\bm{\lambda}^*$; $\bm{u}^*$ and $\bm{\lambda}^*$ are the optimal primal and dual variables, respectively. We take the derivative of the desired acceleration. In a matrix form, the result is given by Eq.(\ref{eq:differentiable-QP-matrix}).
\begin{equation}\label{eq:differentiable-QP-matrix}
\begin{bmatrix} 
\frac{\partial \bm{u}^*}{\partial \hat{a}_{i}(t)} \\ \frac{\partial \lambda^*}{\partial \hat{a}_{i}(t)} 
\end{bmatrix} = K^{-1} 
\begin{bmatrix} 
\mathbb{O} \\ 
\mathbb{D} \left(\bm{\lambda}^*\right) \frac{\partial F\big(\hat{a}_{i}(t)\big)}{\partial \hat{a}_{i}(t)}
\end{bmatrix}
\end{equation}
with 
\begin{equation}
K = 
\begin{bmatrix} 
Q & G^{\mathsf{T}} \\ 
\mathbb{D} \left(\bm{\lambda}^*\right) G & \mathbb{D} \Big( G \bm{u}^* - F\big(\hat{a}_{i}(t)\big)\Big)
\end{bmatrix} 
\end{equation}

Utilizing the partial derivative of the QP layer, we can derive the value of $\frac{\partial u_i(t)}{\partial \hat{a}_{i}(t)}$ in Eq.(\ref{eq:differentiable-QP-loss}) which can be used to facilitate the training of the RL agent.

\section{Results and Analysis}\label{sec:result}

This section conducts simulations to evaluate our proposed platoon-centric safe RL-based GLOSA framework. We first introduce our simulation settings, benchmarks, and training results in Section \ref{sec:simulation-settings}. Then, we evaluate safety, travel efficiency, and energy consumption in Section \ref{sec:testing}. Finally, we analyze the generalizability and scalability of our proposed methods under different driving conditions in Section \ref{sec:scaling}.

\subsection{Simulation Settings} \label{sec:simulation-settings}
\begin{figure}[h]
    \centering
    \makebox[\columnwidth]{\includegraphics[width= \columnwidth]{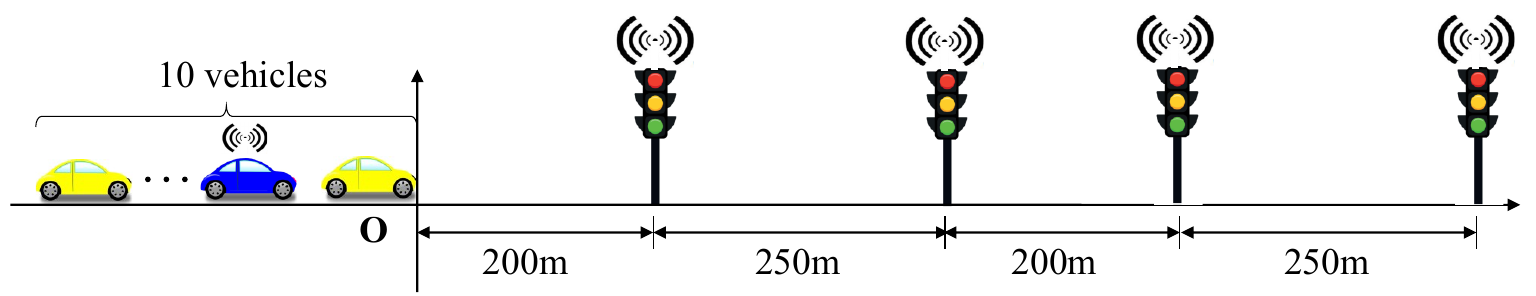}}
    \caption{Simulation settings. A platoon of $10$ vehicles aims to pass through an arterial with $4$ intersections. The coordinates of the leading vehicle are initialized by $0~\text{m}$, and the distances from the leading vehicle to these intersections are $200~\text{m}$, $450~\text{m}$, $650~\text{m}$, and $900~\text{m}$, respectively.}
    \label{fig: experiment}
\end{figure}

As shown in Figure~\ref{fig: experiment}, we consider a typical arterial with four signalized intersections, which implement pre-timed signal controllers optimized to provide bi-directional green waves. The lengths of the incoming leg to the intersections are given between 200$~\rm{m}$ and 250$~\rm{m}$. For each traffic light, the red time and green time, in terms of seconds, are sampled from uniform distributions $\mathbb{U}(5, 10)$ and $\mathbb{U}(10, 12)$, respectively. The offset between two adjacent traffic lights is set to 15 $ \rm{s} $. Note that scenarios with short-cycle signal timing plans are more challenging for mixed-autonomy platoons due to frequent signal switching. Under these settings, the vehicle platoon cannot pass through an intersection within a single green phase, and thus GLOSA is needed to help with the eco-driving of the platoon. Hence, these scenarios increase the testing complexity and allow for a more thorough evaluation of our proposed GLOSA algorithms, as suggested in existing literature \cite{xu2018cooperative, yue2022effects}.

We consider a platoon of 10 vehicles, where the first vehicle has an initial position at $0~\rm{m}$. These vehicles are initialized with spacings following a normal distribution $\mathbb{N}(30, 5)$\,m and with speeds following a normal distribution $\mathbb{N}(13, 3)$\,m/s. The desired minimum headway $h$ is set to be $1.8~\rm{s}$, and the safe stopping distance is $ 2~\rm{m}$. The maximum speed $v^{\rm{max}}$ is set to  $18~\rm{m/s}$. The maximum acceleration and deceleration are set to  $4~\rm{m/s^2}$. The IDM model of HDVs has parameters in Table \ref{tab:IDM-param}. We update the accelerations of CAVs and HDVs every $\Delta T = 0.1~\rm{s}$. We will test our algorithm performance under different CAV penetration rates ranging from $20\%$ to $80\%$.

\begin{table}[h]
    \centering
    \caption{IDM PARAMETERS of HDVs}
    \label{tab:IDM-param}
    \begin{tabular}{c|c}
    \toprule 
        IDM Parameter & Value\\
        \hline
        Maximum acceleration $a_0$ & 1.3 $\rm{m/s^2}$  \\
        Maximum deceleration $b_0$& 4.5 $\rm{m/s^2}$ \\
        Maximum speed $v^{\rm{max}}$& 18 $\rm{m/s}$\\
        Minimum gap distance $s_0$& 2 $\rm{m}$\\
        Desired minimum headway $T$& 1.8 $\rm{s}$\\
        Acceleration exponent $\delta_0$& 4\\
    \bottomrule
    \end{tabular}
\end{table}

The RL agent consists of four MLP layers with $64$, $32$, $16$, and $8$ hidden units to extract features, respectively. The extracted features pass through a $\rm{tanh}$ function to obtain the mean and a $\rm{softplus}$ function to obtain the variance. We specify the activation function to $\rm{sigmoid}$. The learning rates for the actor and critic are set to $10^{-2}$ and $10^{-3}$, respectively. The clip parameter $\epsilon$ in Eq.(\ref{eq:actor-obj}) is set to $0.2$. The weights in reward function $\omega_1$, $\omega_2$, $\omega_3$, and $\omega_4$ are set to $5$, $10$, $5$, and $1$, respectively. The discount factor $\gamma$ is set to $0.9$. We iteratively update the policy 10 times per epoch, computing gradients using 32 batches of transitions each time. The safe RL-based framework is implemented using PyTorch. All RL agents are trained on an NVIDIA GeForce RTX 2070 Super.

\vspace{0.6em}\noindent \textbf{Benchmarks}. We evaluate our proposed GLOSA algorithm by comparing the following benchmarks: 

\begin{itemize}
\item \textbf{RL-Platoon}: our proposed safe RL-based controller.

\item \textbf{PureHDV}: the baseline algorithm where the CAVs adopt the car-following model of HDVs, i.e., IDM. This benchmark represents the current situation to demonstrate the value of performing GLOSA.

\item \textbf{NEcoSA}~\cite{waqas2022correct}: a state-of-the-art algorithm that calculates desired acceleration with a nominal PID controller and adjusts this acceleration to a safe action with CBFs. This benchmark can be considered a variant of our RL-Platoon algorithm without the RL framework. Hence, it is compared with RL-Platoon to evaluate the benefit of incorporating RL for platoon-centric GLOSA.

\item \textbf{HeuSSA}~\cite{guo2019joint, ma2017parsimonious}: a state-of-the-art algorithm that employs a parsimonious shooting heuristic to perform trajectory planning for individual CAVs. It is adapted to mixed-traffic scenarios by integrating a motion predictor for HDVs. Nevertheless, this algorithm is not able to handle the disturbances in the real-time operations of HDVs. Comparing our proposed RL-Platoon with this benchmark allows us to evaluate the benefits of incorporating explicit safety guarantees and platoon-centric reward design.

\textbf{RL-SoftSafe}: a naive RL-based GLOSA system that considers safety via reward shaping without providing any theoretical safety guarantees. This benchmark highlights the limitations of relying solely on reward-driven safety incentives. By comparing this benchmark with RL-Platoon, we demonstrate the superiority of our proposed CBFs for ensuring driving safety over time.
 
\item \textbf{RL-CE}~\cite{ding2024eco}: a state-of-the-art algorithm that leverages an RL agent to control CAVs and develops a feasible speed region to enhance safety. However, this method (i) only enforces weak safety constraints on the current driving speed rather than ensuring the safety of future states as in CBF, and (ii) focuses on individual CAVs rather than the entire platoon. Therefore, by comparing our proposed RL-Platoon with this benchmark, we can evaluate the benefits of providing explicit safety guarantees with CBFs and considering platoon-centric rewards.

\item \textbf{RL-Selfish}: a variant of our proposed safe RL-based controller with CBF-based QP but without considering platoon performance (i.e., vehicle-centric by focusing on individual CAV performance). By comparing this benchmark with RL-CE, we can demonstrate the benefit of providing explicit safety guarantees with CBFs. Moreover, by comparing RL-Platoon with this benchmark, we demonstrate the value of considering platoon-centric reward.

\end{itemize}

\vspace{0.6em}\noindent \textbf{Training Results}. 
We train the RL agents for RL-SoftSafe, RL-CE, RL-Selfish, and RL-Platoon across five independent runs to ensure robustness of our results. These RL agents can converge within 500 epochs, each containing approximately 1200 transitions, for all training scenarios. As each epoch during training takes half a minute, the convergence can be achieved within approximately four hours for each trial. During deployment, the RL agents do not require continuous updates. Therefore, we do not need to compute the derivatives of CBFs, and thus the proposed algorithm supports real-time control, with each action computed less than $0.1~\mathrm{s}$.

\begin{figure}[h]
    \centering
    \includegraphics[width= 0.8\columnwidth]{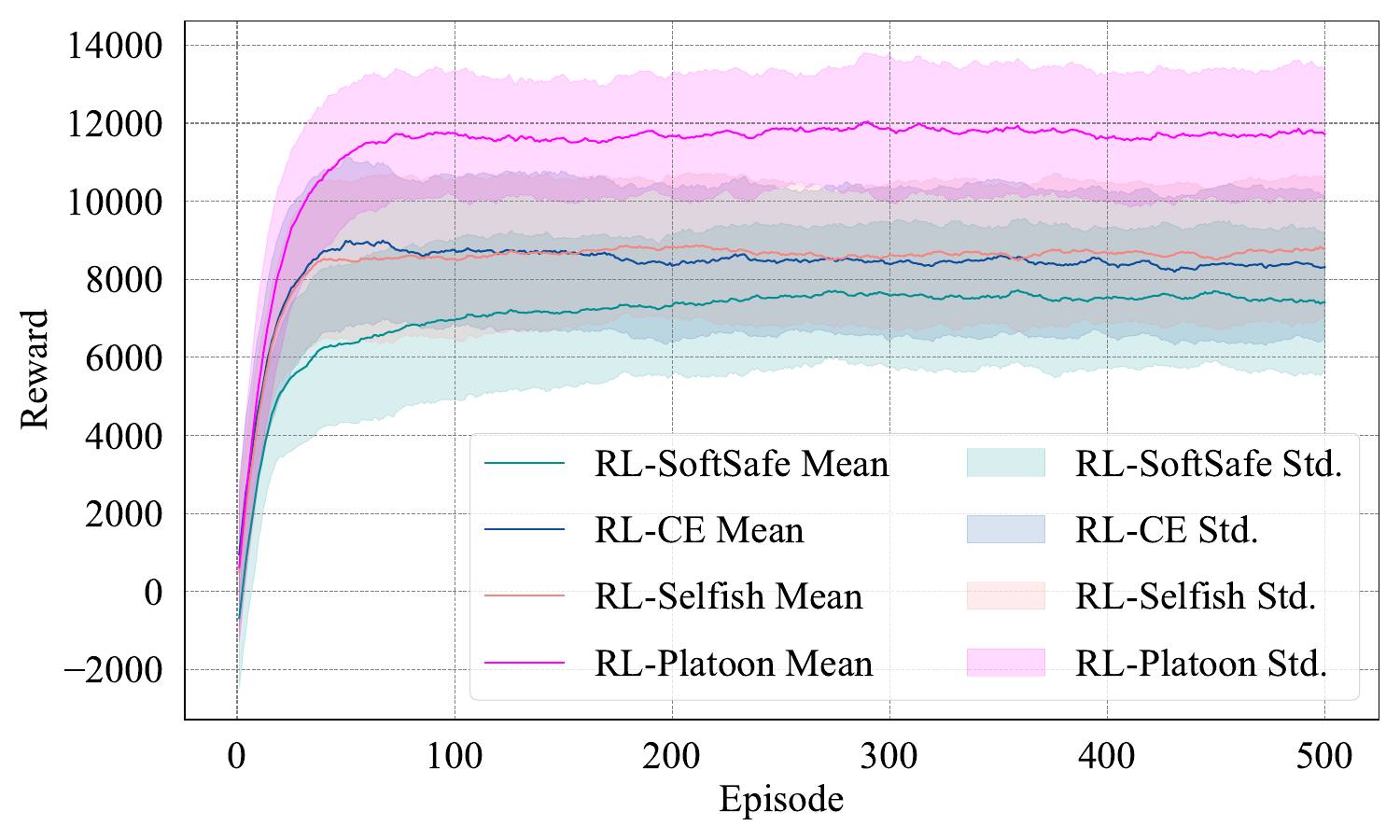}
    \caption{Training Results of RL Benchmarks. The mean and variance are computed across five runs with different random seeds. Without loss of generality, we show the running rewards of the first CAV under a penetration rate of 40\% as an illustration.}
    \label{fig:Training-RL}
\end{figure}

\begin{figure}[h]
    \centering
    \includegraphics[width= 0.8\columnwidth]{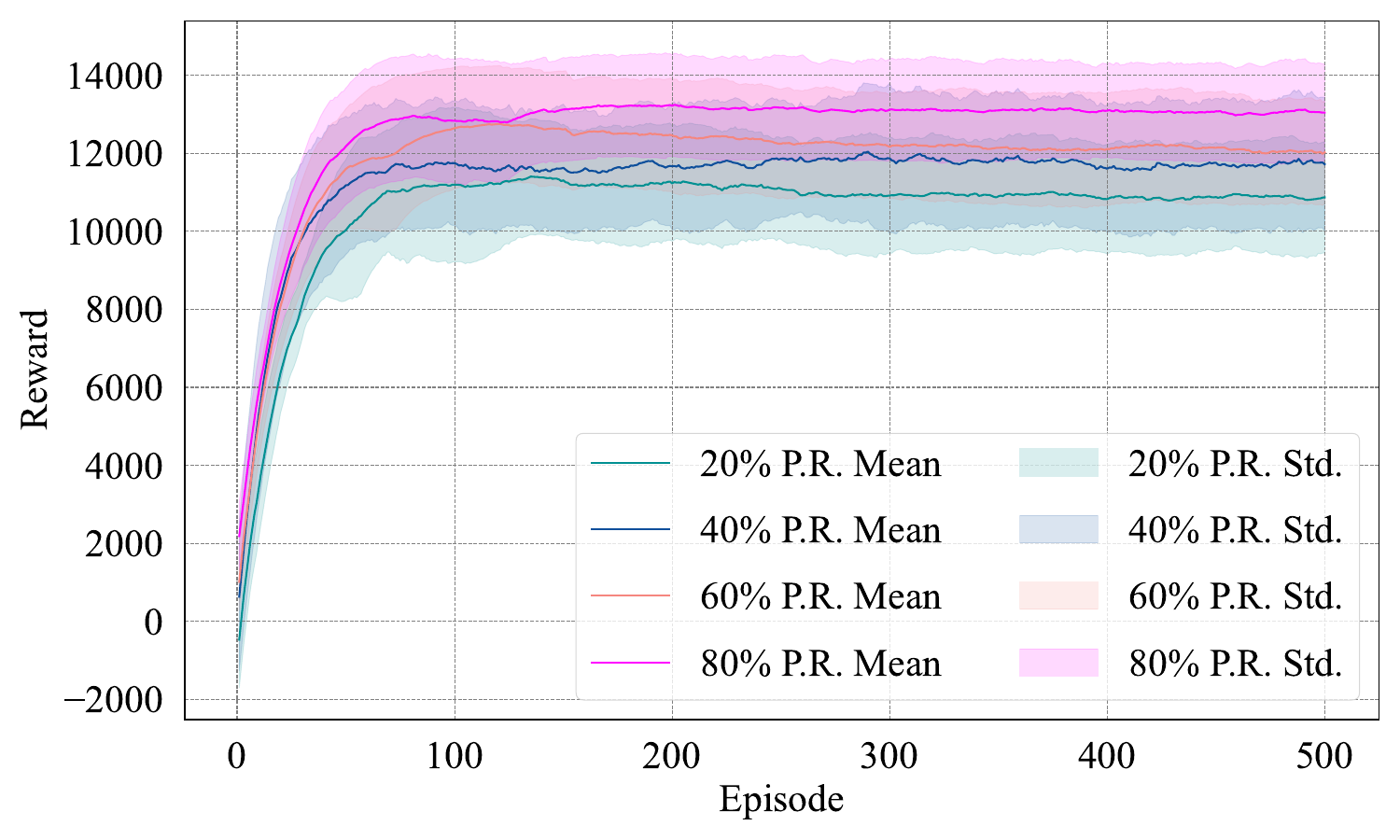}
    \caption{Training Results of RL-Platoon by Different CAV Penetration Rates (P.R.). The training curves consistently converge under different CAV P.R., with the final average reward generally increasing as the CAV P.R. increases.}
    \label{fig:RL-Platoon-PR}
\end{figure}

The training results are shown in Figure \ref{fig:Training-RL} and Figure \ref{fig:RL-Platoon-PR}. From Figure \ref{fig:Training-RL}, we can see that all RL agents converge. Additionally, although RL-Platoon has a more complex reward structure, it converges with similar rates as RL-CE and RL-Selfish, demonstrating the capability of its computational efficiency in real-time control. Note that in Figure \ref{fig:Training-RL}, the reward fluctuates across episodes because the RL agent is trained in diverse scenarios with randomly generated configurations (e.g., signal timing configurations). This diversity is introduced to enhance the generalizability of the trained RL agent and lead to different reward levels across episodes. From Figure \ref{fig:RL-Platoon-PR}, we can see that our proposed MARL-based controller demonstrates stable convergence across all CAV market penetration rates, with a similar convergence rate.

\subsection{Testing results} \label{sec:testing}

\begin{figure*}[h]
    \centering
    \subfigure[PureHDV]{
        \includegraphics[width=0.23\textwidth]{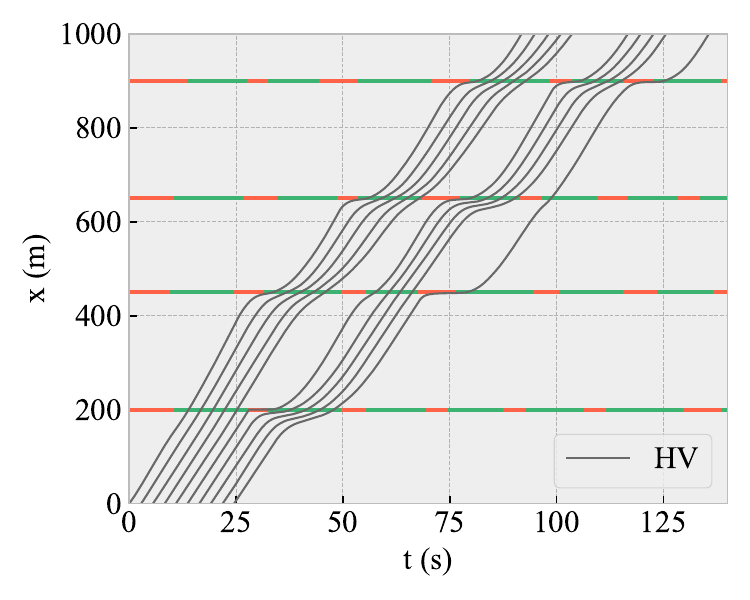}
        \label{fig:Test-PureHVs}         
    }
    \subfigure[NEcoSA]{
        \includegraphics[width=0.23\textwidth]{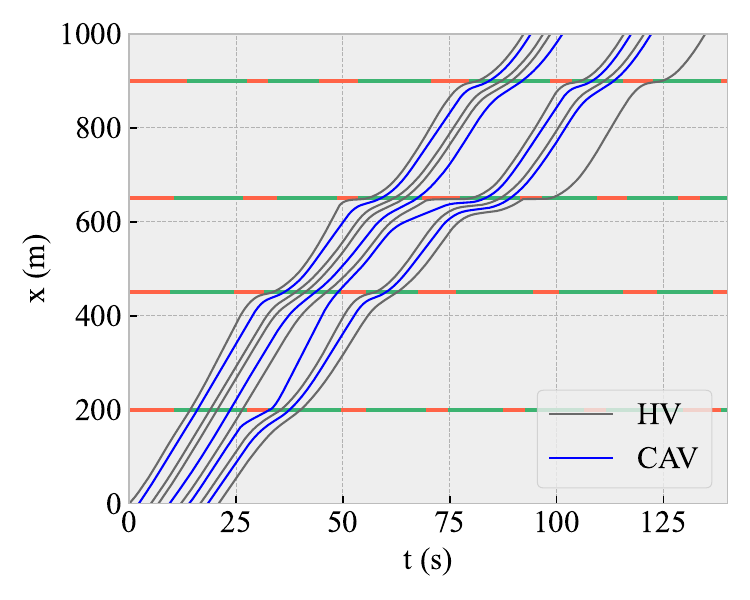}
        \label{fig:Test-NEcoSA}
    }
    \subfigure[HeuSSA]{
        \includegraphics[width=0.23\textwidth]{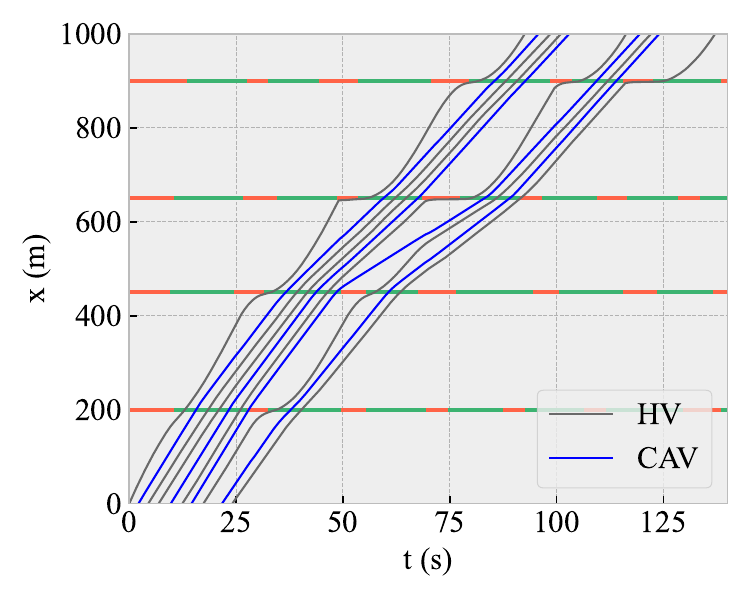}
        \label{fig:Test-HeuSSA}
    }
    \subfigure[RL-SoftSafe]{
        \includegraphics[width=0.23\textwidth]{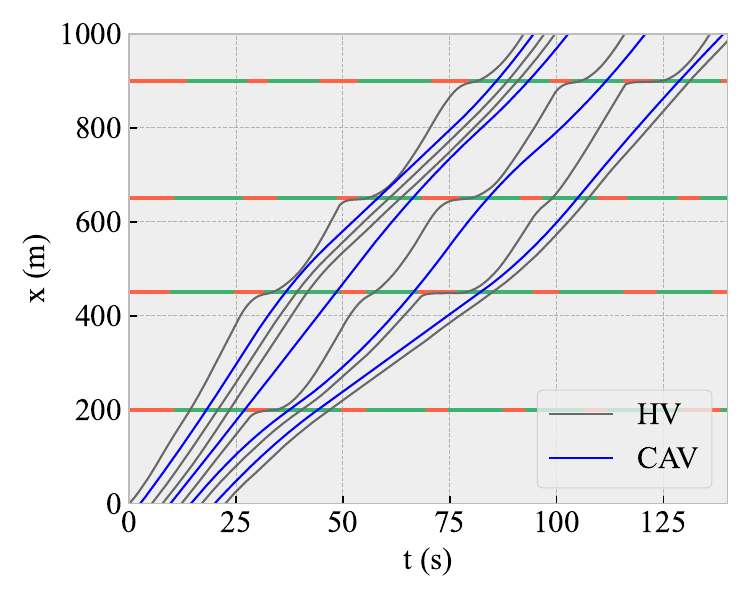}
        \label{fig:Test-RL-SoftSafe}
    }
    \subfigure[RL-CE]{
        \includegraphics[width=0.23\textwidth]{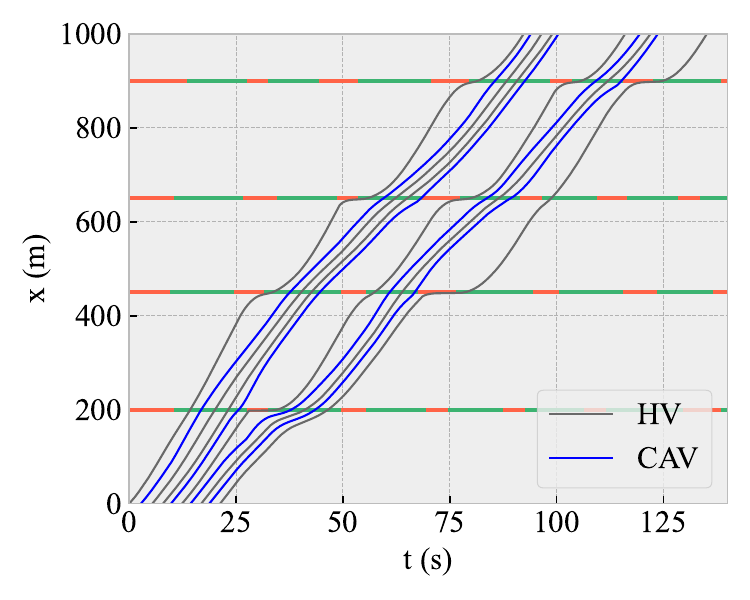}
            \label{fig:Test-RL-CE}
    }
    \subfigure[RL-Selfish]{
        \includegraphics[width=0.23\textwidth]{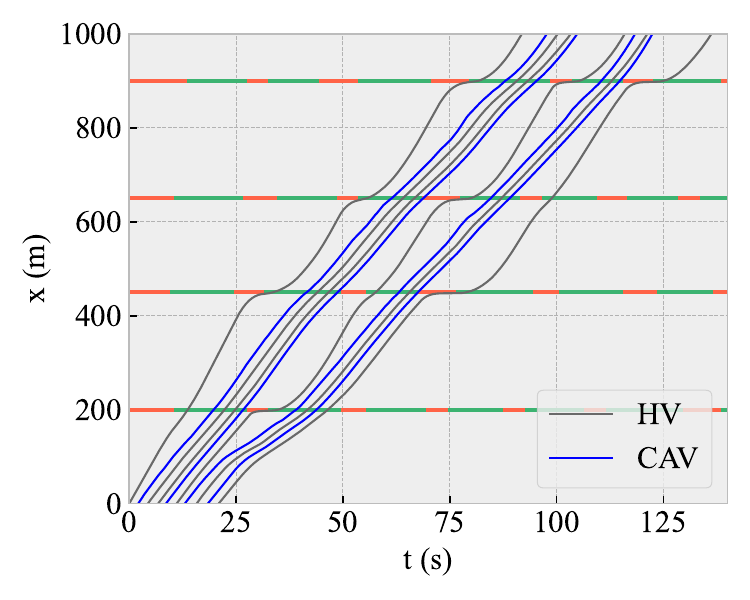}
        \label{fig:Test-RL-Selfish}
    }
    \subfigure[RL-Platoon]{
        \includegraphics[width=0.23\textwidth]{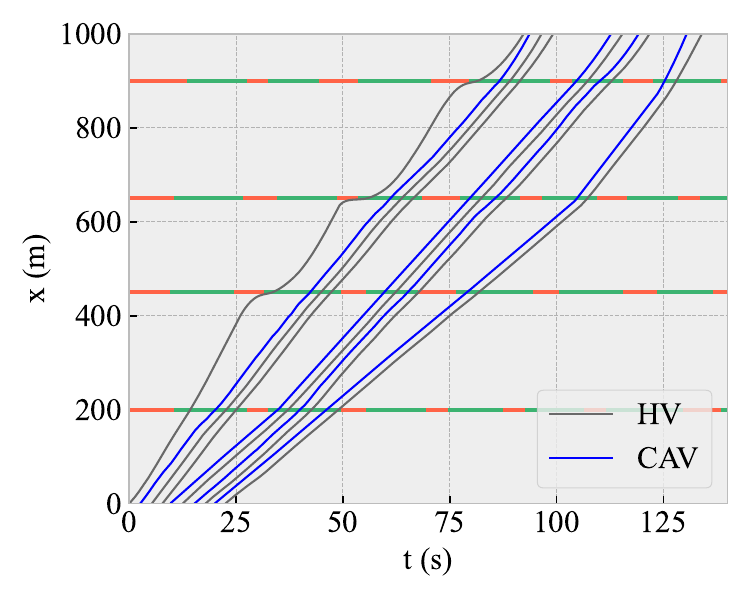}
        \label{fig:Test-RL-Platoon}
    }
    \caption{Time-space diagrams of different algorithms. Compared to other benchmarks, our proposed RL-Platoon enables the entire platoon to pass through the arterial safely without vehicle collisions or red-light violations while facilitating eco-driving.}
    \label{fig:Test-Results}
\end{figure*}

We evaluate the safety, travel efficiency, and energy consumption of the proposed benchmarks under a regular scenario and an extreme scenario. In the regular scenario, no external disturbances are considered, and the environment shares the same parameter distributions as the training environment. The extreme scenarios are constructed based on regular settings, with two additional disturbances: (i) the immediate preceding vehicle of a CAV performs a sudden deceleration of 4 $\rm{m/s^2}$ and (ii) CAVs overestimate the red start time by $2~\rm{s}$, assuming it begins $2~\rm{s}$ later than it actually does. These disturbances are unknown to the CAVs and are intended to evaluate the robustness of the proposed CBFs. Note that the MARL agents are trained only on regular scenarios and have not encountered extreme scenarios. To account for stochasticity, we test these benchmarks in five independent realizations generated with different random seeds. Both the mean values and standard deviations are reported. The testing results are shown in Table~\ref{tab:Test} for the regular scenario and Table~\ref{tab: Test-Disturb} for the extreme scenario.
We also provide time-space diagrams to offer an intuitive illustration of our control performance in Figure~\ref{fig:Test-Results}.  

\begin{table*}[htbp]
    \centering
    \caption{Testing Results in Regular Scenarios.}
    \begin{minipage}{\textwidth}
        \label{tab:Test}
    \centering
    \begin{tabular}{c c|p{1.4 cm} p{1.2 cm} p{0.4 cm}| p{1.2 cm} p{1.2 cm} p{0.4 cm}| p{0.6 cm} p{0.4 cm}|p{1.5 cm} p{0.7 cm} p{0.7 cm}}
    \toprule 
        \multirow{3}{*}{CAV PR.} & \multirow{3}{*}{Algos} & \multicolumn{3}{c|}{Car-Following Safety} &\multicolumn{3}{c|}{Red-Light Safety} & \multicolumn{2}{c|}{Travel Efficiency} & \multicolumn{3}{c}{Energy Consumption}\\
        &  & Dist2PV & TTC & NoC & Dist2TL & T2TL & VoR & AvgV  & T2P & EC & $\Delta$ EC & Imp\\
        &  & ($\rm{m}$) & ($\rm{s}$) & -- & ($\rm{m}$) &($\rm{s}$) & -- &($\rm{m/s}$)  & ($\rm{s}$) & ($\rm{kJ}$) & ($\rm{kJ}$)& (\%)\\
         \hline
         0                      & PureHDV       & $6.9~(\pm 0.6)$  & $2.2~(\pm 0.6)$ & $0$ & $0.0~(\pm 0.0)$ & $0.0~(\pm 0.0)$ & $5$ & $10.8$ & $124.2$ & $9786~(\pm300)$ & -- & -- \\
         \hline
        \multirow{6}{*}{20\%}   & NEcoSA        & $6.6~(\pm 0.6)$  & $2.5~(\pm 0.6)$ & $0$ & $3.6~(\pm 1.9)$ & $0.9~(\pm 0.5)$ & $0$ & $10.7$ & $\bm{123.2}$ & $9941~(\pm 254)$ &  $155$   & $-1.6$ \\
                                & HeuSSA        & $9.6~(\pm 2.3)$ & $3.3~(\pm 0.9)$ & $0$ & $4.3~(\pm 1.7)$ & $1.0~(\pm 0.5)$ & $0$ & $10.8$ & $123.8$ & $8480~(\pm 328)$  & $-1306$ & $13.3$ \\
                                & RL-SoftSafe        & $1.8~(\pm 2.4)$ & $0.3~(\pm 0.5)$ & $6$ & $0.0~(\pm 0.0)$ & $0.0~(\pm 0.0)$ & $13$ & $\bm{11.3}$ & $123.3$ & $8566~(\pm 334)$ & $-1220$ & $12.5$ \\
                                & RL-CE         & $8.4~(\pm 1.4)$ & $3.0~(\pm 0.4)$ & $0$ & $4.3~(\pm 1.7)$ & $1.0~(\pm 0.6)$ & $0$ & $10.8$ & $123.4$ & $8702~(\pm 349)$  & $-1083$ & $11.1$ \\
                                & RL-Selfish    & $10.4~(\pm 2.7)$ & $3.5~(\pm 0.7)$ & $0$ & $4.6~(\pm 1.8)$ & $1.4~(\pm 0.3)$  & $0$ & $10.8$ & $123.7$ & $8127~(\pm 342)$  & $-1659$ & $17.0$ \\
                                & RL-Platoon    & $\bm{11.3~(\pm 3.0)}$ & $\bm{4.0~(\pm 0.7)}$ & $\bm{0}$ & $\bm{7.1~(\pm 0.8)}$ & $2.1~(\pm 1.0)$ & $\bm{0}$ & $10.8$ & $123.4$ & $\bm{7599~(\pm 247)}$ & $\bm{-2187}$  & $\bm{22.3}$ \\
         \hline
        \multirow{6}{*}{40\%}   & NEcoSA        & $6.9~(\pm 0.4)$  & $2.7~(\pm 0.5)$& $0$ & $3.8~(\pm 1.9)$ & $1.0~(\pm 0.6)$ & $0$ & $10.8$ & $\bm{123.0}$ & $9846~(\pm 271)$ & $60$ & $-0.6$ \\
                                & HeuSSA        & $10.3~(\pm 2.6)$ & $3.5~(\pm 0.9)$ & $0$ & $4.5~(\pm 1.8)$ & $1.1~(\pm 0.6)$ & $0$ & $10.9$ & $124.5$ & $8364~(\pm 303)$ & $-1422$ & $14.5$\\
                                & RL-SoftSafe   & $1.9~(\pm 2.7)$ & $0.4~(\pm 0.6)$ & $5$ & $0.4~(\pm 1.0)$ & $0.0~(\pm 0.0)$ & $10$ & $\bm{11.4}$ & $123.9$ &  $8470~(\pm 346)$ & $-1315$ & $13.4$ \\
                                & RL-CE         & $8.7~(\pm 1.3)$ & $3.2~(\pm 0.5)$ & $0$ & $4.6~(\pm 1.8)$ & $1.1~(\pm 0.6)$ & $0$ & $10.9$ & $123.4$ & $8565~(\pm 378)$ & $-1221$ & $12.5$ \\
                                & RL-Selfish    & $10.9~(\pm 2.9)$ & $3.8~(\pm 0.7)$ & $0$ & $5.0~(\pm 1.9)$ & $1.5~(\pm 0.3)$ & $0$ & $10.8$ & $123.4$ & $7947~(\pm 270)$  & $-1839$ & $18.8$ \\
                                & RL-Platoon    & $\bm{11.7~(\pm 2.8)}$ & $\bm{4.3~(\pm 0.8)}$ & $\bm{0}$ & $\bm{8.0~(\pm 1.6)}$ & $\bm{2.4~(\pm 1.1)}$ & $\bm{0}$ & $10.8$ & $125.1$ & $\bm{7431~(\pm 247)}$  & $\bm{-2355}$ & $\bm{24.1}$ \\
         \hline
        \multirow{6}{*}{60\%}   & NEcoSA        & $7.0~(\pm 0.3)$  & $2.8~(\pm 0.5)$ & $0$ & $4.0~(\pm 2.0)$ & $1.1~(\pm 0.6)$ & 0 & $10.9$ & $\bm{123.4}$ & $9726~(\pm 298)$ & $-60$   & $0.6$ \\
                                & HeuSSA        & $10.7~(\pm 2.7)$ & $3.7~(\pm 0.9)$ & $0$ & $5.3~(\pm 0.8)$ & $1.3~(\pm 0.5)$ & $0$ & $11.0$ & $123.4$ & $8108~(\pm 329)$ & $-1677$ & $17.1$ \\
                                & RL-SoftSafe   & $2.1~(\pm 2.8)$ & $0.5~(\pm 0.7)$ & $4$ & $0.9~(\pm 1.3)$ & $0.2~(\pm 0.3)$ & $5$ & $\bm{11.5}$ & $123.7$ & $8227~(\pm 307)$  & $-1559$ & $15.9$ \\
                                & RL-CE         & $9.9~(\pm 2.8)$ & $3.7~(\pm 0.7)$ & $0$ & $6.1~(\pm 1.2)$ & $1.3~(\pm 0.5)$ & $0$ & $10.9$ & $124.9$ &  $8342~(\pm 260)$  & $-1444$ & $14.8$\\
                                & RL-Selfish    & $11.1~(\pm 2.8)$ & $4.1~(\pm 0.7)$ & $0$ &$7.4~(\pm 1.1)$ & $1.8~(\pm 0.9)$ & $0$ & $10.9$ & $124.5$ &  $7792~(\pm 320)$  & $-1994$ & $20.4$\\
                                & RL-Platoon    & $\bm{12.1~(\pm 2.6)}$ & $\bm{4.6~(\pm 0.7)}$ & $\bm{0}$ & $\bm{8.4~(\pm 1.9)}$ & $\bm{2.6~(\pm 1.1)}$ & $\bm{0}$ & $10.9$ & $125.0$ & $\bm{7309~(\pm 232)}$  & $\bm{-2477}$ & $\bm{25.3}$\\
         \hline
        \multirow{6}{*}{80\%}   & NEcoSA        & $7.1~(\pm 0.2)$  & $2.9~(\pm 0.5)$ & $0$ & $4.5~(\pm 2.2)$ & $1.2~(\pm 0.7)$ & $0$ & $11.0$ & $\bm{122.9}$ & $9669~(\pm 296)$ & $-116$  & $1.2$ \\
                                & HeuSSA        & $11.1~(\pm 2.9)$ & $4.0~(\pm 0.9)$ & $0$ & $5.6~(\pm 0.9)$ & $1.4~(\pm 0.5)$ & $0$ & $11.1$ & $124.0$ & $7984~(\pm 401)$ & $-1802$ & $18.4$ \\
                                & RL-SoftSafe        & $2.3~(\pm 3.2)$ & $0.7~(\pm 0.9)$ & $3$ & $1.6~(\pm 1.5)$ & $0.3~(\pm 0.5)$ & $4$ & $\bm{11.7}$ & $123.5$ &  $8092~(\pm 404)$ & $-1694$ & $17.3$ \\
                                & RL-CE         & $10.2~(\pm 2.7)$ & $3.8~(\pm 0.6)$ & $0$ & $6.4~(\pm 1.2)$ & $1.4~(\pm 0.5)$ & $0$ & $11.1$ & $124.1$ & $8132~(\pm 357)$  & $-1654$ & $16.9$ \\
                                & RL-Selfish    & $11.4~(\pm 2.7)$ & $4.3~(\pm 0.7)$ & $0$ & $7.9~(\pm 1.3)$ & $2.1~(\pm 1.0)$ & $0$ & $11.0$ & $124.6$ & $7665~(\pm 238)$  & $-2121$ & $21.7$ \\
                                & RL-Platoon    & $\bm{12.2~(\pm 2.5)}$ & $\bm{4.8~(\pm 0.7)}$ & $\bm{0}$ & $\bm{8.6~(\pm 2.2)}$ & $\bm{2.7~(\pm 1.2)}$ & $\bm{0}$ & $11.0$ & $124.4$ & $\bm{7195~(\pm 277)}$  & $\bm{-2591}$ & $\bm{26.5}$ \\
    \bottomrule
    \end{tabular}
    \end{minipage}
\end{table*}

\begin{table*}[htbp]
    \centering
    \caption{Testing Results in More Extreme Conditions with Larger Disturbances.}
    \begin{minipage}{\textwidth}
        \label{tab: Test-Disturb}
    \centering
    \begin{tabular}{c c|p{1.4 cm} p{1.2 cm} p{0.4 cm}| p{1.2 cm} p{1.2 cm} p{0.4 cm}| p{0.6 cm} p{0.4 cm}|p{1.5 cm} p{0.7 cm} p{0.7 cm}}
    \toprule 
        \multirow{3}{*}{CAV PR.} & \multirow{3}{*}{Algos} & \multicolumn{3}{c|}{Car-Following Safety} &\multicolumn{3}{c|}{Red-Light Safety} & \multicolumn{2}{c|}{Travel Efficiency} & \multicolumn{3}{c}{Energy Consumption}\\
        &  & Dist2PV & TTC & NoC & Dist2TL & T2TL & VoR & AvgV  & T2P & EC & $\Delta$EC & Imp\\
        &  & ($\rm{m}$) & ($\rm{s}$) & -- & ($\rm{m}$) &($\rm{s}$) & -- &($\rm{m/s}$)  & ($\rm{s}$) & ($\rm{kJ}$) & ($\rm{kJ}$)& (\%)\\
         \hline
         0                      & PureHDV       & $1.0~(\pm 2.1)$ & $0.1~(\pm 0.2)$ & $19$ & $0.0~(\pm 0.0)$ & $0.0~(\pm 0.0)$ & $19$ & $12.2$ & $122.6$ & $9735~(\pm 320)$ & -- & -- \\
         \hline
        \multirow{6}{*}{20\%}   & NEcoSA        & $5.3~(\pm 0.9)$  & $1.4~(\pm 1.0)$ & $0$ & $2.9~(\pm 1.4)$ & $0.5~(\pm 0.4)$ & $0$ & $12.0$ & $122.6$ & $9917~(\pm 298)$ &  $182$   & $-1.9$ \\
                                & HeuSSA        & $0.0~(\pm 0.0)$  & $0.0~(\pm 0.0)$  & $18$ & $0.0~(\pm 0.0)$ & $0.0~(\pm 1.0)$ & $17$ & $\bm{12.2}$ & $\bm{122.0}$ & $8489~(\pm 294)$  & $-1246$ & $12.8$ \\
                                & RL-SoftSafe        & $0.0~(\pm 0.0)$ & $0.0~(\pm 0.0)$ & $19$ & $0.0~(\pm 0.0)$ & $0.0~(\pm 0.0)$ & $17$ & $11.9$ & $122.1$ & $8446~(\pm 189)$ & $-1289$ & $13.2$ \\
                                & RL-CE         & $0.0~(\pm 0.0)$ & $0.0~(\pm 0.0)$ & $13$ & $0.0~(\pm 0.0)$ & $0.0~(\pm 0.0)$ & $7$ & $12.0$ & $122.4$ & $8733~(\pm 370)$  & $-1002$ & $10.3$ \\
                                & RL-Selfish    & $7.9~(\pm 2.6)$ & $2.0~(\pm 1.2)$ & $0$ & $4.1~(\pm 1.5)$ & $1.0~(\pm 0.2)$ & $0$ & $11.3$ & $122.4$ & $8142~(\pm 401)$  & $-1593$ & $16.4$ \\
                                & RL-Platoon    & $\bm{8.8~(\pm 2.6)}$ & $\bm{2.5~(\pm 1.1)}$ & $\bm{0}$ & $\bm{6.5~(\pm 1.3)}$ & $\bm{1.4~(\pm 0.8)}$ & $\bm{0}$ & $11.8$ & $122.1$ & $\bm{7613~(\pm 314)}$ & $\bm{-2122}$ & $\bm{21.8}$\\
         \hline
        \multirow{6}{*}{40\%}   & NEcoSA        & $5.7~(\pm 0.9)$  & $1.7~(\pm 1.0)$ & $0$ & $3.2~(\pm 1.4)$ & $0.6~(\pm 0.5)$ & $0$ & $11.9$ & $\bm{121.6}$ & $9821~(\pm 302)$ & $86$    & $-0.9$ \\
                                & HeuSSA        & $0.0~(\pm 0.0)$ & $0.0~(\pm 0.0)$ & $16$ & $0.0~(\pm 0.0)$ & $0.0~(\pm 0.0)$ & $17$ & $11.8$ & $122.5$ & $8393~(\pm 269)$ & $-1342$ & $13.8$\\
                                & RL-SoftSafe        & $0.0~(\pm 0.0)$ & $0.0~(\pm 0.0)$ & $17$ & $0.0~(\pm 0.0)$ & $0.0~(\pm 0.0)$ & $15$ &  $\bm{12.2}$ & $122.6$ & $8376~(\pm 198)$ & $-1359$ & $14.0$\\
                                & RL-CE         & $0.0~(\pm 0.0)$ & $0.0~(\pm 0.0)$ & $12$ & $0.4~(\pm 0.9)$ & $0.1~(\pm 0.2)$ & $5$ & $11.6$ & $122.1$ & $8627~(\pm 335)$  & $-1108$ & $11.4$ \\
                                & RL-Selfish    & $8.3~(\pm 2.5)$ & $3.0~(\pm 1.2)$ & $0$ & $4.7~(\pm 1.7)$ & $1.2~(\pm 0.2)$ & $0$ & $11.8$ & $122.2$ & $7938~(\pm 318)$  & $-1798$ & $18.5$ \\
                                & RL-Platoon    & $\bm{9.4~(\pm 2.4)}$ & $\bm{3.5~(\pm 1.0)}$ & $\bm{0}$ & $\bm{6.8~(\pm 1.4)}$ & $\bm{1.6~(\pm 0.8)}$ & $\bm{0}$ & $11.6$ & $123.6$ & $\bm{7416~(\pm 238)}$  & $\bm{-2319}$ & $\bm{23.8}$ \\
         \hline
        \multirow{6}{*}{60\%}   & NEcoSA        & $6.0~(\pm 0.8)$  & $1.9~(\pm 1.0)$ & $0$ & $3.4~(\pm 1.5)$ & $0.8~(\pm 0.5)$ & $0$ & $11.9$ & $\bm{122.1}$ & $9720~(\pm 290)$ & $-15$   & $0.2$ \\
                                & HeuSSA        & $0.0~(\pm 0.0)$ & $0.0~(\pm 0.0)$ & $14$ & $0.0~(\pm 0.0)$ & $0.0~(\pm 0.0)$ & $13$ & $12.2$ & $122.2$ & $8099~(\pm 186)$ & $-1636$ & $16.8$ \\
                                & RL-SoftSafe        & $0.0~(\pm 0.0)$ & $0.0~(\pm 0.0)$ & $14$ & $0.0~(\pm 0.0)$ & $0.0~(\pm 0.0)$ & $11$ & $\bm{12.3}$ & $122.2$ &  $8255~(\pm 143)$ & $-1480$ & $15.2$ \\
                                & RL-CE         & $0.6~(\pm 1.4)$ &  $0.1~(\pm 0.1)$  & $7$ &  $1.9~(\pm 1.8)$  & $0.4~(\pm 0.4)$ & $2$ & $11.8$ & $122.6$ & $8336 (\pm 161)$ & $-1400$ & $14.4$ \\
                                & RL-Selfish    &  $8.7~(\pm 2.6)$  & $3.2~(\pm 1.3)$ & $0$ & $6.3~(\pm 1.1)$ & $1.6~(\pm 0.9)$ & $0$ & $12.2$ & $122.3$ & $7795~(\pm 361)$  & $-1940$ & $19.9$\\
                                & RL-Platoon    & $\bm{9.6~(\pm 2.3)}$ & $\bm{3.8~(\pm 1.0)}$ & $\bm{0}$ & $\bm{7.1~(\pm 1.3)}$ & $\bm{1.8~(\pm 1.0)}$ & $\bm{0}$ & $12.2$ & $122.4$ & $\bm{7284~(\pm 224)}$  & $\bm{-2451}$ & $\bm{25.2}$\\
         \hline
        \multirow{6}{*}{80\%}   & NEcoSA        & $6.2~(\pm 0.8)$  & $2.0~(\pm 1.0)$ & $0$ & $3.7~(\pm 1.6)$ & $0.9~(\pm 0.6)$ & $0$ & $11.7$ & $\bm{122.0}$ & $9661~(\pm 302)$ & $-74$  & $0.8$ \\
                                & HeuSSA        & $1.0~(\pm 2.1)$ & $0.1~(\pm 0.3)$ & $9$ & $0.0~(\pm 0.0)$ & $0.0~(\pm 0.0)$ & $10$ & $12.2$ & $122.7$ & $7962~(\pm 227)$ & $-1773$ & $18.2$ \\
                                & RL-SoftSafe        & $0.0~(\pm 0.0)$ & $0.0~(\pm 0.0)$ & $11$ & $0.0~(\pm 0.0)$ & $0.0~(\pm 0.0)$ & $9$ & $11.8$ & $122.2$ & $8111~(\pm 221)$ & $-1624$ & $16.7$ \\
                                & RL-CE         & $1.6~(\pm 2.2)$ & $0.2~(\pm 0.3)$ & $4$ & $3.0~(\pm 1.7)$ & $0.6~(\pm 0.4)$ & $1$ & $12.0$ & $122.5$ & $8153~(\pm 226)$ & $-1582$ & $16.2$ \\
                                & RL-Selfish    & $9.1~(\pm 2.5)$ & $3.4~(\pm 1.2)$ & $0$ & $6.7~(\pm 1.0)$ & $1.9~(\pm 1.1)$ & $0$ & $12.2$ & $122.5$ & $7586~(\pm 298)$  & $-2149$ & $22.1$ \\
                                & RL-Platoon    & $\bm{9.8~(\pm 2.3)}$ & $\bm{4.0~(\pm 1.1)}$ & $\bm{0}$ & $\bm{7.2~(\pm 1.3)}$ &$\bm{2.0~(\pm 1.2)}$ & $\bm{0}$ & $\bm{12.3}$ & $122.3$ & $\bm{7190~(\pm 240)}$  & $\bm{-2545}$ & $\bm{26.1}$ \\
    \bottomrule
    \end{tabular}
    \end{minipage}
\end{table*}

\begin{table*}[ht]
\begin{minipage}{\textwidth}
\centering
\caption{Energy Consumption Evaluation under Different Platoon Sizes}
\label{tab:Scaling}
\begin{tabular}{c c|c c c c||c c|c c c c}
\hline
\multirow{2}{*}{Size} & \multirow{2}{*}{Algos} & EC & Std. & $\Delta$ EC & Imp & \multirow{2}{*}{Size} & \multirow{2}{*}{Algos} & EC & Std. & $\Delta$ EC & Imp \\
&  & ($\rm{kJ}$) & -- & ($\rm{kJ}$) & (\%) & & &($\rm{kJ}$) & -- & ($\rm{kJ}$) & (\%)\\
\hline
\multirow{7}{*}{$$4$$} & PureHDV & $$4203$$ & $$293$$ & -- & -- & \multirow{7}{*}{$$10$$} & PureHDV & $$9786$$ & $$300$$ & -- & -- \\
 & NEcoSA & $$4327$$ & $$255$$ & $$124$$ & $$-3.0$$ & & NEcoSA & $$9846$$ & $$271$$ & $$60$$ & $$-0.6$$ \\
 & HeuSSA & $$3604$$ & $$299$$ & $$-599$$ & $$14.3$$ & & HeuSSA & $$8364$$ & $$303$$ & $$-1422$$ & $$14.5$$ \\
 & RL-SoftSafe & $$3671$$ & $$342$$ & $$-532$$ & $$12.7$$ & & RL-SoftSafe & $$8470$$ & $$346$$ & $$-1316$$ & $$13.4$$ \\
 & RL-CE & $$3812$$ & $$366$$ & $$-391$$ & $$9.3$$ & & RL-CE & $$8565$$ & $$378$$ & $$-1221$$ & $$12.5$$ \\
 & RL-Selfish & $$3144$$ & $$268$$ & $$-1059$$ & $$25.2$$ & & RL-Selfish & $$7947$$ & $$270$$ & $$-1839$$ & $$18.8$$ \\
 & RL-Platoon & $\bm{2815}$ & $\bm{241}$ & $\bm{-1388}$ & $\bm{33.0}$ & & RL-Platoon & $\bm{7431}$ & $\bm{247}$ & $\bm{-2355}$ & $\bm{24.1}$ \\
\hline
\multirow{7}{*}{$$6$$} & PureHDV & $$6166$$ & $$297$$ & -- & -- & \multirow{7}{*}{$$12$$} & PureHDV & $$11414$$ & $$303$$ & -- & -- \\
 & NEcoSA & $$6264$$ & $$261$$ & $$98$$ & $$-1.6$$ & & NEcoSA & $$11390$$ & $$274$$ & $$-24$$ & $$0.2$$ \\
 & HeuSSA & $$5411$$ & $$300$$ & $$-755$$ & $$12.2$$ & & HeuSSA & $$9766$$ & $$304$$ & $$-1648$$ & $$14.4$$ \\
 & RL-SoftSafe & $$5494$$ & $$343$$ & $$-672$$ & $$10.9$$ & & RL-SoftSafe & $$9854$$ & $$348$$ & $$-1560$$ & $$13.7$$ \\
 & RL-CE & $$5677$$ & $$370$$ & $$-489$$ & $$7.9$$ & & RL-CE & $$9913$$ & $$379$$ & $$-1501$$ & $$13.2$$ \\
 & RL-Selfish & $$4931$$ & $$269$$ & $$-1235$$ & $$20.0$$ & & RL-Selfish & $$9232$$ & $$271$$ & $$-2182$$ & $$19.1$$ \\
 & RL-Platoon & $\bm{4423}$ & $\bm{244}$ & $\bm{-1743}$ & $\bm{28.3}$ & & RL-Platoon & $\bm{8761}$ & $\bm{250}$ & $\bm{-2653}$ & $\bm{23.2}$ \\
\hline
\multirow{7}{*}{$$8$$} & PureHDV & $$8073$$ & $$299$$ & -- & -- & \multirow{7}{*}{$$14$$} & PureHDV & $$12871$$ & $$307$$ & -- & -- \\
 & NEcoSA & $$8152$$ & $$268$$ & $$79$$ & $$-1.0$$ & & NEcoSA & $$12811$$ & $$275$$ & $$-60$$ & $$0.5$$ \\
 & HeuSSA & $$7011$$ & $$301$$ & $$-1062$$ & $$13.2$$ & & HeuSSA & $$11072$$ & $$305$$ & $$-1799$$ & $$14.0$$ \\
 & RL-SoftSafe & $$7125$$ & $$344$$ & $$-948$$ & $$11.7$$ & & RL-SoftSafe & $$11171$$ & $$350$$ & $$-1700$$ & $$13.2$$ \\
 & RL-CE & $$7329$$ & $$373$$ & $$-744$$ & $$9.2$$ & & RL-CE & $$11242$$ & $$380$$ & $$-1629$$ & $$12.7$$ \\
 & RL-Selfish & $$6511$$ & $$269$$ & $$-1562$$ & $$19.3$$ & & RL-Selfish & $$10471$$ & $$272$$ & $$-2400$$ & $$18.6$$ \\
 & RL-Platoon & $\bm{5981}$ & $\bm{245}$ & $\bm{-2092}$ & $\bm{25.9}$ & & RL-Platoon & $\bm{10014}$ & $\bm{252}$ & $\bm{-2857}$ & $\bm{22.2}$ \\
\hline
\end{tabular}
\end{minipage}
\end{table*}

\begin{table*}[htbp]
    \centering
    \caption{Testing Results under Different Communication Ranges in Regular Scenarios.}
    \begin{minipage}{\textwidth}
        \label{tab:CR_impact}
    \centering
    \begin{tabular}{c c|p{1.4 cm} p{1.2 cm} p{0.4 cm}| p{1.2 cm} p{1.2 cm} p{0.4 cm}| p{0.6 cm} p{0.4 cm}|p{1.5 cm} p{0.7 cm} p{0.7 cm}}
    \toprule 
        \multirow{3}{*}{CR.($\rm{m}$)} & \multirow{3}{*}{Algos} & \multicolumn{3}{c|}{Car-Following Safety} &\multicolumn{3}{c|}{Red-Light Safety} & \multicolumn{2}{c|}{Travel Efficiency} & \multicolumn{3}{c}{Energy Consumption}\\
        &  & Dist2PV & TTC & NoC & Dist2TL & T2TL & VoR & AvgV  & T2P & EC & $\Delta$ EC & Imp\\
         &  & ($\rm{m}$) & ($\rm{s}$) & -- & ($\rm{m}$) &($\rm{s}$) & -- &($\rm{m/s}$)  & ($\rm{s}$) & ($\rm{kJ}$) & ($\rm{kJ}$)& (\%)\\
         \hline
         0                      & PureHDV       & $6.9~(\pm 0.6)$  & $2.2~(\pm 0.6)$ & $0$ & $0.0~(\pm 0.0)$ & $0.0~(\pm 0.0)$ & $5$ & $10.8$ & $124.2$ & $9786~(\pm300)$ & -- & -- \\
         \hline
        \multirow{6}{*}{100}   & NEcoSA        & $3.3~(\pm 1.8)$  & $1.9~(\pm 0.7)$ & $0$ & $2.0~(\pm 0.9)$ & $0.5~(\pm 0.2)$ & $0$ & $10.7$ & $\bm{123.0}$ & $10311~(\pm 267)$ &  $525$   & $-5.4$ \\
                                & HeuSSA        & $6.4~(\pm 2.3)$ & $2.1~(\pm 0.6)$ & $0$ & $2.2~(\pm 0.9)$ & $0.8~(\pm 0.3)$ & $0$ & $11.0$ & $124.6$ & $8821~(\pm 278)$  & $-965$ & $9.9$ \\
                                & RL-SoftSafe        & $0.0~(\pm 0.0)$ & $0.0~(\pm 0.0)$ & $7$ & $0.0~(\pm 0.0)$ & $0.0~(\pm 0.0)$ & $11$ & $\bm{11.4}$ & $123.9$ & $8949~(\pm 297)$ & $-837$ & $8.6$ \\
                                & RL-CE         & $5.1~(\pm 1.7)$ & $2.0~(\pm 0.6)$ & $0$ & $2.5~(\pm 1.1)$ & $0.9~(\pm 0.3)$ & $0$ & $10.9$ & $123.4$ & $9118~(\pm 265)$  & $-668$ & $6.8$ \\
                                & RL-Selfish    & $7.0~(\pm 2.2)$ & $2.1~(\pm 0.7)$ & $0$ & $2.8~(\pm 1.2)$ & $1.1~(\pm 0.3)$  & $0$ & $10.7$ & $123.5$ & $8758~(\pm 275)$  & $-1028$ & $10.5$ \\
                                & RL-Platoon    & $\bm{8.2~(\pm 1.8)}$ & $\bm{2.7~(\pm 0.7)}$ & $\bm{0}$ & $\bm{5.3~(\pm 1.5)}$ & $1.6~(\pm 0.3)$ & $\bm{0}$ & $10.8$ & $125.0$ & $\bm{8512~(\pm 287)}$ & $\bm{-1273}$  & $\bm{13.0}$ \\
         \hline
        \multirow{6}{*}{150}   & NEcoSA        & $5.8~(\pm 2.1)$  & $2.2~(\pm 0.6)$& $0$ & $3.0~(\pm 1.0)$ & $0.6~(\pm 0.4)$ & $0$ & $10.9$ & $\bm{123.0}$ & $10244~(\pm 243)$ & $438$ & $-4.5$ \\
                                & HeuSSA        & $8.9~(\pm 1.8)$ & $3.1~(\pm 0.8)$ & $0$ & $3.1~(\pm 0.9)$ & $0.8~(\pm 0.4)$ & $0$ & $10.8$ & $124.6$ & $8564~(\pm 248)$ & $-1222$ & $12.5$\\
                                & RL-SoftSafe   & $0.0~(\pm 0.0)$ & $0.0~(\pm 0.0)$ & $7$ & $0.0~(\pm 0.0)$ & $0.0~(\pm 0.0)$ & $11$ & $\bm{11.4}$ & $123.7$ &  $8638~(\pm 234)$ & $-1148$ & $11.7$ \\
                                & RL-CE         & $7.7~(\pm 2.1)$ & $2.7~(\pm 1.0)$ & $0$ & $3.3~(\pm 0.8)$ & $1.0~(\pm 0.5)$ & $0$ & $10.9$ & $123.5$ & $8768~(\pm 254)$ & $-1018$ & $10.4$ \\
                                & RL-Selfish    & $9.4~(\pm 2.2)$ & $3.0~(\pm 0.9)$ & $0$ & $3.8~(\pm 0.8)$ & $1.1~(\pm 0.6)$ & $0$ & $10.8$ & $123.4$ & $8393~(\pm 279)$  & $-1393$ & $14.2$ \\
                                & RL-Platoon    & $\bm{10.5~(\pm 2.3)}$ & $\bm{3.6~(\pm 0.8)}$ & $\bm{0}$ & $\bm{6.7~(\pm 1.1)}$ & $\bm{1.8~(\pm 0.8)}$ & $\bm{0}$ & $11.0$ & $125.2$ & $\bm{7893~(\pm 287)}$  & $\bm{-1893}$ & $\bm{19.3}$ \\
         \hline
        \multirow{6}{*}{200}   & NEcoSA        & $6.4~(\pm 1.9)$  & $2.5~(\pm 0.8)$ & $0$ & $3.4~(\pm 1.7)$ & $0.8~(\pm 0.4)$ & 0 & $10.9$ & $\bm{122.9}$ & $10051~(\pm 354)$ & $265$   & $-2.7$ \\
                                & HeuSSA        & $9.4~(\pm 2.3)$ & $3.4~(\pm 0.5)$ & $0$ & $4.0~(\pm 2.0)$ & $1.1~(\pm 0.6)$ & $0$ & $10.9$ & $124.5$ & $8488~(\pm 243)$ & $-1298$ & $13.3$ \\
                                & RL-SoftSafe   & $1.4~(\pm 2.8)$ & $0.3~(\pm 0.7)$ & $5$ & $0.0~(\pm 0.0)$ & $0.0~(\pm 0.0)$ & $11$ & $\bm{11.4}$ & $123.9$ & $8569~(\pm 313)$  & $-1217$ & $12.4$ \\
                                & RL-CE         & $8.2~(\pm 2.3)$ & $3.0~(\pm 0.8)$ & $0$ & $4.3~(\pm 1.2)$ & $1.1~(\pm 0.5)$ & $0$ & $10.8$ & $123.5$ &  $8626~(\pm 233)$  & $-1160$ & $11.9$\\
                                & RL-Selfish    & $10.2~(\pm 2.2)$ & $3.4~(\pm 0.6)$ & $0$ &$4.5~(\pm 1.2)$ & $1.4~(\pm 0.7)$ & $0$ & $10.9$ & $123.4$ &  $8174~(\pm 325)$  & $-1612$ & $16.5$\\
                                & RL-Platoon    & $\bm{11.2~(\pm 2.1)}$ & $\bm{3.9~(\pm 0.4)}$ & $\bm{0}$ & $\bm{7.6~(\pm 1.5)}$ & $\bm{2.3~(\pm 0.6)}$ & $\bm{0}$ & $11.0$ & $125.0$ & $\bm{7615~(\pm 228)}$  & $\bm{-2171}$ & $\bm{22.2}$\\
         \hline
        \multirow{6}{*}{250}   & NEcoSA        & $6.9~(\pm 0.4)$  & $2.7~(\pm 0.5)$& $0$ & $3.8~(\pm 1.9)$ & $1.0~(\pm 0.6)$ & $0$ & $10.8$ & $\bm{123.0}$ & $9846~(\pm 271)$ & $60$ & $-0.6$ \\
                                & HeuSSA        & $10.3~(\pm 2.6)$ & $3.5~(\pm 0.9)$ & $0$ & $4.5~(\pm 1.8)$ & $1.1~(\pm 0.6)$ & $0$ & $10.9$ & $124.5$ & $8364~(\pm 303)$ & $-1422$ & $14.5$\\
                                & RL-SoftSafe   & $1.9~(\pm 2.7)$ & $0.4~(\pm 0.6)$ & $5$ & $0.4~(\pm 1.0)$ & $0.0~(\pm 0.0)$ & $10$ & $\bm{11.4}$ & $123.9$ &  $8470~(\pm 346)$ & $-1315$ & $13.4$ \\
                                & RL-CE         & $8.7~(\pm 1.3)$ & $3.2~(\pm 0.5)$ & $0$ & $4.6~(\pm 1.8)$ & $1.1~(\pm 0.6)$ & $0$ & $10.9$ & $123.4$ & $8565~(\pm 378)$ & $-1221$ & $12.5$ \\
                                & RL-Selfish    & $10.9~(\pm 2.9)$ & $3.8~(\pm 0.7)$ & $0$ & $5.0~(\pm 1.9)$ & $1.5~(\pm 0.3)$ & $0$ & $10.8$ & $123.4$ & $7947~(\pm 270)$  & $-1839$ & $18.8$ \\
                                & RL-Platoon    & $\bm{11.7~(\pm 2.8)}$ & $\bm{4.3~(\pm 0.8)}$ & $\bm{0}$ & $\bm{8.0~(\pm 1.6)}$ & $\bm{2.4~(\pm 1.1)}$ & $\bm{0}$ & $10.8$ & $125.1$ & $\bm{7431~(\pm 247)}$  & $\bm{-2355}$ & $\bm{24.1}$ \\
         \hline
        \multirow{6}{*}{300}   & NEcoSA        & $7.2~(\pm 1.3)$  & $3.1~(\pm 0.9)$ & $0$ & $4.1~(\pm 0.7)$ & $1.3~(\pm 0.6)$ & $0$ & $10.8$ & $\bm{123.1}$ & $9659~(\pm 318)$ & $-127$   & $1.3$ \\
                                & HeuSSA        & $10.7~(\pm 1.5)$ & $3.9~(\pm 0.8)$ & $0$ & $4.9~(\pm 1.5)$ & $1.5~(\pm 0.6)$ & $0$ & $10.9$ & $124.5$ & $8238~(\pm 374)$ & $-1548$ & $15.8$ \\
                                & RL-SoftSafe   & $2.2~(\pm 2.8)$ & $0.9~(\pm 1.2)$ & $3$ & $0.7~(\pm 1.4)$ & $0.1~(\pm 0.2)$ & $8$ & $\bm{11.5}$ & $123.7$ & $8460~(\pm 288)$  & $-1326$ & $13.5$ \\
                                & RL-CE         & $9.8~(\pm 1.6)$ & $3.8~(\pm 0.9)$ & $0$ & $5.1~(\pm 0.8)$ & $1.7~(\pm 0.5)$ & $0$ & $10.8$ & $123.3$ &  $8519~(\pm 223)$  & $-1267$ & $12.9$\\
                                & RL-Selfish    & $11.2~(\pm 1.9)$ & $4.0~(\pm 0.9)$ & $0$ &$5.6~(\pm 0.7)$ & $2.1~(\pm 0.6)$ & $0$ & $10.8$ & $123.4$ &  $7854~(\pm 372)$  & $-1932$ & $19.7$\\
                                & RL-Platoon    & $\bm{11.7~(\pm 2.3)}$ & $\bm{4.4~(\pm 0.8)}$ & $\bm{0}$ & $\bm{8.7~(\pm 1.5)}$ & $\bm{3.2~(\pm 0.3)}$ & $\bm{0}$ & $10.7$ & $125.1$ & $\bm{7359~(\pm 276)}$  & $\bm{-2427}$ & $\bm{24.8}$\\
    \bottomrule
    \end{tabular}
    \end{minipage}
\end{table*}

\subsubsection{Safety}
The car-following safety is evaluated using the following metrics: (i) the minimum distance to the preceding vehicle (Dist2PV), (ii) the minimum time to collision (TTC), and (iii) total number of collisions (NoC). The red-light safety is evaluated via (i) the minimum distance to the next traffic light if the signal indication of this traffic light is red (Dist2TL), (ii) the time required to reach this traffic light if the vehicle were to continue at its current speed (T2TL), and (iii) total red-light violations (VoR). 

We make the following observations from Tables~\ref{tab:Test} and \ref{tab: Test-Disturb} on the safety performance. First, by comparing the performance of RL-SoftSafe and RL-CE with RL-Platoon, we can see that the CBF-based approach can improve all safety criteria, especially in scenarios with low CAV penetration rates. This demonstrates the value of CBF in enhancing safety over time in learning-based controllers. Second, we observe that by performing platoon-centric control, RL-Platoon can generally improve safety performance compared to RL-Selfish. This is because platoon-centric methods can avoid sudden deceleration of CAV following vehicles, especially for HDVs, as demonstrated by Figure~\ref{fig:Test-Results}. Third, unlike RL-CE and HeuSSA, RL-Selfish and RL-Platoon achieve zero NoC and VoR under disturbances, demonstrating the robustness of the CBFs under disturbances as illustrated in Table \ref{tab: Test-Disturb}. Fourth, with the increasing CAV penetration rates, we observe an increase in all these safety indicators. This illustrates that with the massive deployment of CAVs, our proposed methods can be more effective in enhancing safety. 

\subsubsection{Travel Efficiency and Energy Consumption}
We evaluate travel efficiency using (i) the average speed (AvgV) and (ii) the time required for the entire platoon to pass the last intersection (T2P). We observe from Tables~\ref{tab:Test} and \ref{tab: Test-Disturb} that these benchmarks have similar performance in travel efficiency, i.e., AvgV and T2P. Moreover, with the increase in CAV penetration rates, T2P tends to increase. One possible explanation is that CAVs may choose to pass the intersection during the next green phase to avoid stopping, as illustrated in Figure \ref{fig:Test-Results}.
 
Next, we evaluate the energy consumption through (i) the total energy consumption of the entire platoon (EC), (ii) the change in energy consumption compared to PureHDV ($\Delta$EC), and (iii) the percentage improvement in energy consumption computed by ER divided by EC (Imp). According to Tables \ref{tab:Test} and \ref{tab: Test-Disturb}, we make the following remarks regarding energy efficiency. First, we can see that our proposed RL-platoon significantly outperforms state-of-the-art (e.g., HeuSSA and RL-CE) in EC, $\Delta$ EC, and Imp. This illustrates that platoon-centric methods can further improve energy efficiency. The underlying reason for higher energy efficiency is that platoon-centric methods facilitate a smoother trajectory not only for the CAV itself but also for the following vehicles, as demonstrated in Figure \ref{fig:Test-Results}. Second, since the energy consumption of RL-CE and RL-Selfish is similar, the embedded CBF-based safety module does not lead to an increase in energy consumption. This enables us to expand our methods to address more complicated safety challenges. Third, as the penetration rates increase, the RL-platoon reduces energy consumption to a greater extent. This illustrates how our proposed methods become more effective with the spread and advancements of CAVs.

\subsection{Generalizability and Scalability} \label{sec:scaling}
We further investigate the performance of our proposed RL-Platoon under different driving conditions. Specifically, we examine (i) its generalizability across varying communication ranges and (ii) its scalability with respect to different platoon sizes. Simulations are conducted under a CAV penetration rate of 40\% within the regular scenario.

\subsubsection{Generalizability} We evaluate the performance of these benchmarks under different communication ranges from $100~\rm{m}$ to $300~\rm{m}$, summarized in Table \ref{tab:CR_impact}.

From Table~\ref{tab:CR_impact}, we draw the following key observations. First, as the communication range (CR) of CAVs increases, both car-following safety and red-light safety improve consistently. This is because a longer CR enables CAVs to anticipate traffic conditions earlier and respond proactively, thereby reducing the need for abrupt deceleration. Second, regardless of the CR, our proposed RL-Platoon consistently outperforms all baselines across all safety metrics. This demonstrates the robustness and effectiveness of our CBF-based QP controller in maintaining driving safety under varying CR conditions. Third, RL-Platoon achieves the lowest energy consumption among all benchmarks, highlighting its efficiency and justifying its deployment even under limited CR. Fourth, although increasing the CR contributes to improved safety, its impact on energy efficiency becomes marginal beyond a certain point. This is because with sufficiently large CR, the resulting platoon trajectories converge to a similar pattern, limiting further gains. These findings collectively validate the effectiveness and robustness of our proposed RL-Platoon across different CR settings.

\subsubsection{Scalability} 
We evaluate the energy consumption of these benchmarks under different platoon sizes ranging from $4$ to $14$, as travel efficiency remains similar and safety is generally preserved. The result is shown in Table~\ref{tab:Scaling}. 

From Table~\ref{tab:Scaling}, we make the following observations. 
First, the proposed RL-Platoon consistently achieves the lowest energy consumption among all benchmarks, demonstrating the scalability of our platoon-centric reward design across different platoon sizes. Second, the low standard deviations indicate that RL-Platoon effectively stabilizes the acceleration profiles of surrounding HDVs through CAV coordination. Our proposed platoon-centric reward design not only improves the energy efficiency of CAVs but may also induce positive impacts on the energy consumption of following vehicles. Third, the absolute value of energy savings improves with larger platoon sizes, primarily due to the increasing number of CAVs within each platoon. Fourth, the relative improvement reduces with larger platoon sizes. This is because as more CAVs are deployed, the total number of HDVs in the platoon also increases, potentially undermining the benefits brought by the platoon-centric control strategy. Therefore, we recommend prioritizing the development and deployment of CAVs rather than merely increasing the number of vehicles in each platoon.

To sum up, we conclude that our proposed platoon-centric, safe RL-based GLOSA system outperforms state-of-the-art benchmarks in balancing platoon energy consumption and travel efficiency while ensuring both car-following safety and red-light safety. 

\section{Conclusion and Future work}\label{sec:conclusion}

In this paper, we developed a platoon-centric, safe RL-based GLOSA framework that enables vehicle platoons to safely approach, split, and pass through intersections. We formulate this problem as MARL and leverage CTDE to train the RL agent. Moreover, we leverage a CBF-based QP to ensure car-following safety and red light safety. The gradient information of this QP is passed to the actor to facilitate the training of RL through a differentiable QP layer. The simulation results illustrate that our proposed method outperforms state-of-the-art methods in balancing platoon driving safety, travel efficiency, and energy consumption. Sensitivity analysis demonstrates that our proposed RL-Platoon maintains strong generalizability across varying communication ranges and exhibits good scalability with respect to different platoon sizes.

This paper opens several directions for future research. First, this framework can be extended to consider more complex safety constraints, e.g., the safety of lane-changing or pedestrian crossing. Second, we can consider heterogeneous driving behaviors that lead to different vehicle dynamics. Third, we can extend our proposed framework to incorporate lane-changing maneuvers, which may rely on Vehicle-to-Vehicle (V2V) communication to achieve higher-level vehicle cooperation. Fourth, we would like to extend the method to cope with environmental uncertainties, e.g., signal timing uncertainty. Fifth, it would be interesting to transfer this speed control framework to other transportation scenarios, e.g., merges and diverges. In addition to theoretical investigations, it would be valuable to conduct on-site experiments to validate the effectiveness of our proposed GLOSA system under real-world conditions. During deployment, practical challenges such as communication latency, localization errors, sensor uncertainty, and hardware constraints may arise and require further investigation.

\ifCLASSOPTIONcaptionsoff
  \newpage
\fi

\bibliography{reference}
\newpage
\bibliographystyle{IEEEtran}
\begin{IEEEbiography}
[{\includegraphics[width=1in,height=1.25in,clip,keepaspectratio]{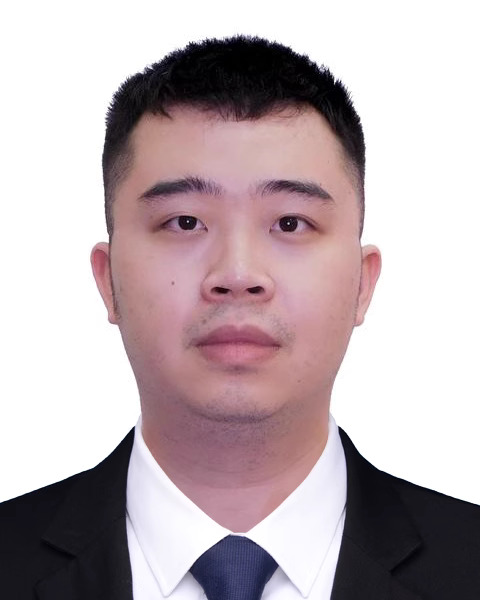}}]
{Ruining Yang}
received his B.Eng. degree from Southeast University, Nanjing, China, in 2023, and his M.Eng. degree from the National University of Singapore in 2025. He is currently pursuing his Ph.D. degree with the School of Civil and Environmental Engineering at the Georgia Institute of Technology, Atlanta, GA, USA. His research interests include intelligent transportation systems, traffic control, and connected and autonomous vehicles.
\end{IEEEbiography}

\begin{IEEEbiography}
[{\includegraphics[width=1in,height=1.25in, clip,keepaspectratio]{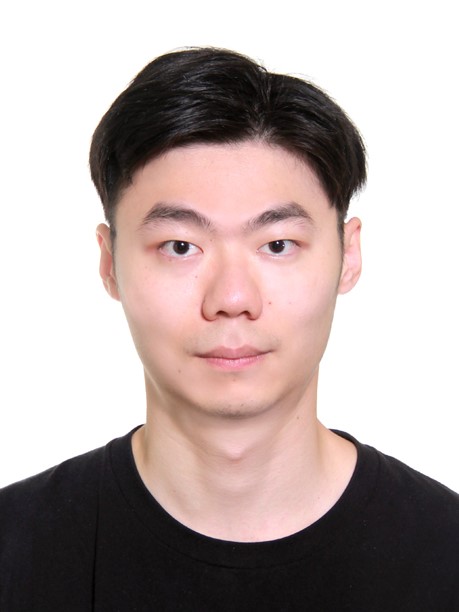}}]{Jingyuan Zhou} receives the B.Eng. degree in Electronic Information Science and Technology from Sun Yat-sen University, Guangzhou, China, in 2022. He is currently working towards a Ph.D. degree with the National University of Singapore. His research interests include safety-critical control and privacy computing of mixed-autonomy traffic.
\end{IEEEbiography}

\begin{IEEEbiography}
[{\includegraphics[width=1in,height=1.25in,clip,keepaspectratio]{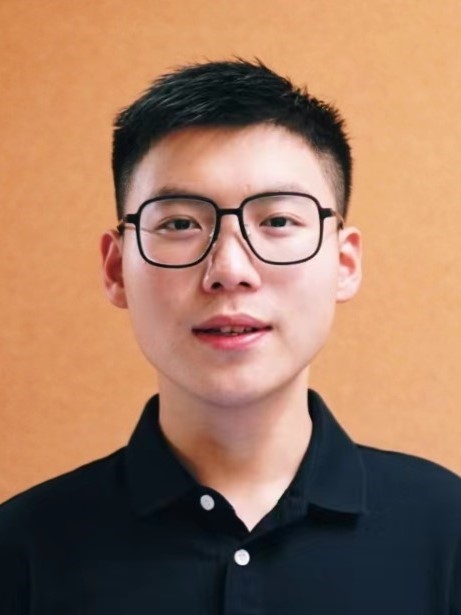}}]
{Qiqing Wang} received his B.Eng. degree in the School of Transportation Science and Engineering from Harbin Institute of Technology, Harbin, China, in 2022. He is currently working towards a Ph.D. degree with the National University of Singapore. His research interests include intelligent transportation systems, traffic control, data collaboration, and data privacy.
\end{IEEEbiography}

\begin{IEEEbiography}
[{\includegraphics[width=1in,height=1.25in,clip,keepaspectratio]{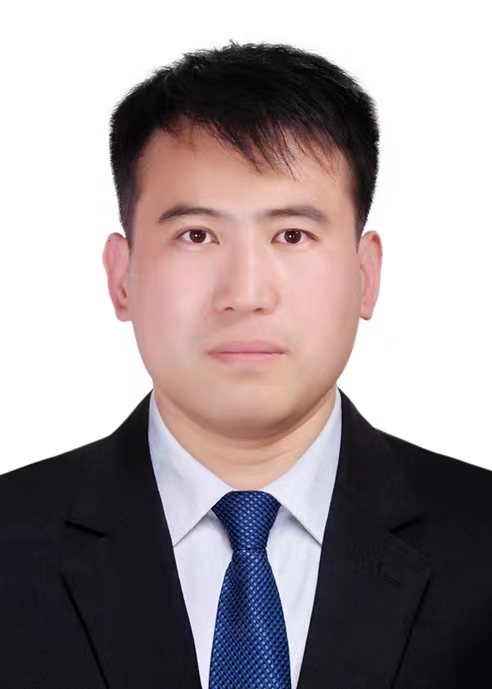}}]{Jinhao Liang}
received the B.S. degree from School of Mechanical Engineering, Nanjing University of Science and Technology, Nanjing, China, in 2017, and Ph.D. degree from School of Mechanical Engineering, Southeast University, Nanjing, China, in 2022. Now he is a Research Fellow with Department of Civil and Environmental Engineering, National University of Singapore. His research interests include vehicle dynamics and control, autonomous vehicles, and vehicle safety assistance system.
\end{IEEEbiography}
\vspace{11pt}

\begin{IEEEbiography}[{\includegraphics[width=1in,height=1.25in,clip,keepaspectratio]{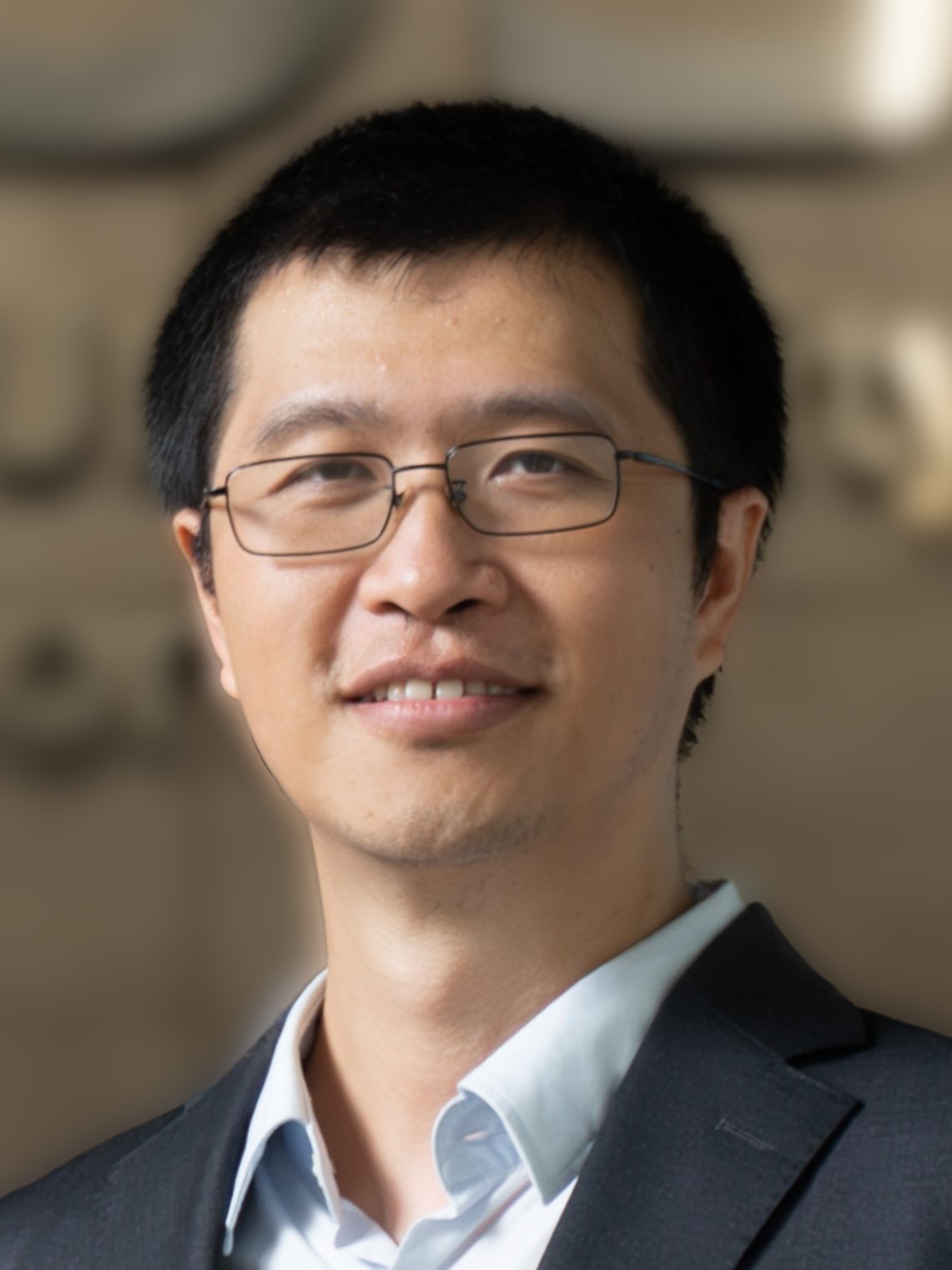}}]{Kaidi Yang}
is an Assistant Professor in the Department of Civil and Environmental Engineering at the National University of Singapore. Prior to this, he was a postdoctoral researcher with the Autonomous Systems Lab at Stanford University. He obtained a PhD degree from ETH Zurich and M.Sc. and B.Eng. degrees from Tsinghua University. His main research interest is the operation of future mobility systems enabled by connected and automated vehicles (CAVs) and shared mobility.
\end{IEEEbiography}
\end{document}